\documentstyle[eepic,epsf,12pt]{article}
\def\gmaru{\hbox{$\hskip0.3em\raisebox{-.0ex}{$g$}
 \raisebox{1.1ex}{\hskip-0.4em$\scriptscriptstyle{\circ}$\hskip0.1em}$}}
\def\etamaru{\hbox{$\hskip0.3em\raisebox{-.0ex}{$\eta$}
 \raisebox{1.1ex}{\hskip-0.4em$\scriptscriptstyle{\circ}$\hskip0.1em}$}}
\def\emaru{\hbox{$\hskip0.3em\raisebox{-.0ex}{$e$}
 \raisebox{1.1ex}{\hskip-0.4em$\scriptscriptstyle{\circ}$\hskip0.1em}$}}

\def\Fmaru{\hbox{$\hskip0.3em\raisebox{-.0ex}{$F$}
 \raisebox{1.8ex}{\hskip-0.4em$\scriptscriptstyle{\circ}$\hskip0.1em}$}}
\def\Dmaru{\hbox{$\hskip0.3em\raisebox{-.0ex}{${\cal D}$}
 \raisebox{1.8ex}{\hskip-0.4em$\scriptscriptstyle{\circ}$\hskip0.1em}$}}
\def\bigBox{\raisebox{-0.1em}{\large\boldmath $\Box$}}
\def\Dslash{\not\hskip -0.7ex{D}}

\newfont{\bg}{cmr10 scaled\magstep4}

\newcommand{\bigzerou}{\smash{\lower1.7ex\hbox{\bg 0}}}

\newcounter{aleq}

\newcommand{\CS}[1]{\!\!\!&{#1}&\!\!\!}

\newcommand{\texthalf}{\textstyle{1\over 2}}
\newcommand{\texthalves}{\textstyle{3\over 2}}

\newcommand{\AdS}{{\it AdS}_{3}}
\newcommand{\sh}{\textstyle{{1\over 2}}}
\newcommand{\sth}{\textstyle{{3\over 2}}}
\newcommand{\LS}{\bigl[ }
\newcommand{\RS}{\bigr] }
\newcommand{\ls}{\left[ }
\newcommand{\rs}{\right] }
\newcommand{\lp}{\left.}
\newcommand{\rp}{\right.}
\newcommand{\NN}{\nonumber}

\newcommand{\ket}[1]{|{#1}\rangle}

\newcommand{\BE}{\begin{equation}}
\newcommand{\EE}{\end{equation}}
\newcommand{\BEA}{\begin{eqnarray}}
\newcommand{\EEA}{\end{eqnarray}}
\newcommand{\BEAN}{\begin{eqnarray*}}
\newcommand{\EEAN}{\end{eqnarray*}}

\newcommand{\JMP}[3]{{\rm J. Math. Phys.} {\bf #1} {(#2)} {#3}}

\renewcommand{\theequation}{\arabic{section}.\arabic{equation}}
\begin{document}
\baselineskip=18pt
%
\begin{titlepage}
\begin{center}
\vspace*{0.5cm}
{\Large{\bf $D=5$ Simple Supergravity on $AdS_{3}\times S^{2}$ and\\
       $N=4$ Superconformal Field Theory}}
\vskip 1.0cm
{\large Akira Fujii\footnote{\tt fujii@tanashi.kek.jp}, 
Ryuji Kemmoku\footnote{\tt kemmoku@tanashi.kek.jp} 
and Shun'ya Mizoguchi\footnote{\tt mizoguch@tanashi.kek.jp}}
\vskip 0.8cm
{\small {\sl Institute of Particle and Nuclear Studies\\
High Energy Accelerator Research Organization (KEK)\\
Tanashi, Tokyo 188-8501, Japan}}
\vskip 0.2cm
\end{center}
\begin{abstract}
We study the Kaluza-Klein spectrum of $D=5$ simple supergravity on $S^2$ 
with special interest in the relation to a two-dimensional $N=4$ 
superconformal field theory. The spectrum is obtained around 
the maximally supersymmetric Freund-Rubin-like background 
$AdS_3\times S^2$ by closely following the well-known techniques 
developed in $D=11$ supergravity. All the vector excitations 
turn out to be ``(anti-)self-dual'', having only one dynamical degree of 
freedom. The representation theory for the Lie superalgebra 
$SU(1,1|2)$ is developed by means of the oscillator method. 
We calculate the conformal weight of 
the boundary operator by estimating the asymptotic behavior of the wave 
function for each Kaluza-Klein mode. All the towers of particles are 
shown to fall into four infinite series of chiral primary 
representations of $SU(1,1|2)\times SL(2,{\bf R})$ (direct product), or 
$OSp(2,2|2;-1)\cong SU(1,1|2)\times SL(2,{\bf R})$ (semi-direct product). 
\end{abstract}
\end{titlepage}
\setcounter{footnote}{0}
%
%
\section{Introduction}
For some time now, we have recognized that string/M theory on an 
Anti-de Sitter space times some compact space is intimately related to 
the conformal field theory on the boundary. The original conjecture 
on this relation \cite{Maldacena:1997re} was clarified in the subsequent 
studies \cite{Witten:1998qj,Gubser:1998bc} and further explored by a number of authors. See 
\cite{Aharony:1999ti} for a recent review and an exhaustive reference.

In this paper we study the correspondence of $D=5$ simple supergravity 
compactified on $S^2$ to a two-dimensional $N=4$ superconformal field 
theory on the boundary of the Anti-de Sitter space $AdS_3$.  
$D=5$ simple supergravity has a BPS solitonic string solution with 
a similar world-sheet structure to that of M5-brane 
\cite{Gibbons:1995vm,Mizoguchi:1998wv}
in particular, (4,0) unbroken world-sheet 
supersymmetry and near-horizon geometry $AdS_3\times S^2$. 
Thus one naturally expects the correspondence between $D=5$ simple 
supergravity on $AdS_3\times S^2$ and a two-dimensional chiral $N=4$ 
superconformal field theory at long distances. 
See 
\cite{Balasubramanian:1998ee}
- \cite{Rahmfeld:1998zn} 
for other works on $AdS_3$.

Let us comment on the relation of the setting in this paper to 
M-theory compactifications. M5-brane wrapped around a holomorphic 
4-cycle of a Calabi-Yau space represents at low energies a black string 
in $D=5$, $N=2$ (``${\cal N}=1$'') supergravity with worldsheet 
(4,0) supersymmetry \cite{Maldacena:1997de}. $D=5$ black string with 
the $AdS_3\times S^2$ near-horizon geometry also arises as the $T^6$ 
compactification of the orthogonal intersection of three M5-branes 
\cite{Kallosh:1997qw,Boonstra:1997dy} 
with leaving the common string and three overall 
transverse dimensions uncompactified. The Kaluza-Klein spectrum of 
D=11 supergravity in such a geometry was already obtained in refs.
\cite{Larsen:1998xm,deBoer:1998ip}. 
Although there is some overlap between their and our results,
we would like to stress a few new contributions made in this paper:

First, we present the complete detail of the derivation, including 
various gauge conditions and the harmonic expansions. This enables 
us to discuss the nature of the pure-gauge doubleton-like modes. 
Secondly, we clarify the existence of the central element in the 
oscillator representation of $SU(1,1|2)$, to which no analogue 
exists in $SU(1,1|M)$ for $M\geq 3$ \cite{Gunaydin:1986fe}. We address 
the issue of the possibility that the symmetry group of the spectrum 
can be $OSp(2,2|2;\alpha\rightarrow -1)$. (See below in this section.)
Finally, we emphasize the striking parallelism between $D=5$ 
simple supergravity on $AdS_3\times S^2$ and $D=11$ supergravity 
on $AdS_7\times S^4$ \cite{Pilch:1984xy}
- \cite{Gunaydin:1985wc}, which is buried in a number of moduli fields of 
the $D=5$, $N=8$ supergravity. Note that no Calabi-Yau or orbifold 
compactification is known to yield $D=5$ simple supergravity, although 
it may be realized as a consistent truncation. It seems hard to orbifold 
away the dilaton \cite{Dabholkar:1998kv}, but we mention an attempt 
to construct some no-moduli theories \cite{Dine:1997ji}. In view of the 
similarity between $D=11$ and $D=5$ simple supergravity theories 
\cite{Cr,Chamseddine:1980sp,Gibbons:1995vm,Mizoguchi:1998wv}, 
it is tempting to suspect that the latter might also be a low-energy 
theory of something fundamental. This issue is beyond the scope of the 
present article.

Since the solitonic string may be regarded as an analogue of M5-brane,
it is useful to gain insights from the 
$AdS_7\times S^4$ compactification of $D=11$ 
supergravity.  It is well-known in the $AdS_7\times S^4$ 
case that the Kaluza-Klein spectrum falls into the representation of 
the supergroup $OSp(6,2|4)$ \cite{Gunaydin:1985wc}. 
Hence the first guess is that our supergroup might be $OSp(2,2|2)$, 
since it has the maximal subgroup $O(2,2)\times USp(2)$ 
$\cong SL(2,{\bf R})_L\times SL(2,{\bf R})_R\times SU(2)$, 
which is nothing but the isometry group of $AdS_3\times S^2$. 
However, it turns out to be unsuited for 
our expectation; the generators of $SL(2,{\bf R})_L$ do 
not decouple but are present in the anticommutators of right-moving 
supercharges. 
$OSp(2,2|2)$ has, on the other hand, infinitely many ``cousins'' 
who all contain $SL(2,{\bf R})_L\times 
SL(2,{\bf R})_R\times SU(2)$ bosonic subgroup, namely 
the one-parameter family of supergroups 
$OSp(2,2|2;\alpha)$.\footnote{or, (an appropriate real form of) 
$D(2,1|\alpha$).}
$OSp(2,2|2)$ itself can be written as $OSp(2,2|2;-\frac12)$. A remarkable   
property of $OSp(2,2|2;\alpha)$ is that it has a decoupling limit 
$\alpha\rightarrow -1$, in which it factorizes into a {\em semi-direct 
product} $SU(1,1|2)_R\times SL(2,{\bf R})_L$. 
Since the Lie algebra of $SU(1,1|2)$ is the special subalgebra of the 
two-dimensional minimal $N=4$ superconformal algebra, 
$OSp(2,2|2;\alpha\!\rightarrow\! -1)$ can be a candidate for our 
supergroup. Note that this is not a direct product, because the decoupled 
$SL(2,{\bf R})_L$ {\em does} act as a rotation and a scale 
transformation on the supercharges.

To obtain the mass spectrum, we will closely follow the techniques 
which were already developed in 80's in the studies of the $11=7+4$ 
\cite{Pilch:1984xy}
- \cite{Gunaydin:1985wc} or $4+7$ compactification 
\cite{Biran:1984iy}
- \cite{Gunaydin:1986tc}. 
(See \cite{Duff:1986hr} for a review and further references therein.) 
In spite of the similarity, some of our results are new and even 
surprising. For example, we will show 
that {\em all} the massive vector excitations are 
``{\em (anti-)self-dual}'' in the sense that the gauge potential is 
proportional to its three-dimensional curl. 
This property should be compared with the ``(anti-)self-dual'' three-form 
gauge field in $D=7$ gauged supergravity \cite{Townsend:1984xs}. 
Another unexpected 
observation is that the zero modes on the solitonic string do {\em not} 
fall into the ``doubleton'' representation, but to what may be called 
``{\em quarteton}'' representation!; although ultra-short doubleton can 
also be constructed in our case, it appears to play no role in the story. 
The quarteton is made up of {\em two pairs} of super-oscillators, 
and turns out to be a pure gauge mode just like singleton or doubleton 
\cite{NiSeTa}
in M theory. The ``massless'' supergravity matter multiplet comes
next, with four pairs of super-oscillators.

The remainder of this paper is organized as follows. In sect.2, we will 
briefly review some basic facts about $D\!=\!5$ simple supergravity 
and $AdS_3$. In sect.3, the Kaluza-Klein mass spectrum around the 
maximally supersymmetric vacuum $AdS_3\times S^2$ is derived. 
The highest weight representations of $SU(1,1|2)$ are studied in sect.4 
by using the oscillator method. In sect.5, we will calculate 
the conformal weight of the boundary operators. In sect.6, we will show 
how each of the Kaluza-Klein modes falls into an $SU(1,1|2)\times
SL(2,{\bf R})$ chiral primary multiplet. The final section summarizes
our results.

Before closing this section, we remark on our convention of the
spacetime signature. The ``mostly positive'' ([$-+\cdots +$]) 
metric is usually used in the modern literature of string theory,
whereas the ``mostly negative'' ([$+-\cdots -$]) one was commonly used 
in the Kaluza-Klein supergravity literature. To facilitate comparison 
with other literature, we adopt [$+-\cdots -$] in sect.3, 
where the Kaluza-Klein spectrum is derived, while we use 
[$-+\cdots +$] in all the other sections. 
%
%
\setcounter{equation}{0}
\section{$D\!=\!5$ Simple Supergravity and $AdS_3$}
\subsection{Solitonic String in $D\!=\!5$ Simple Supergravity}

It has been shown \cite{Gibbons:1995vm,Mizoguchi:1998wv} that 
the magnetic BPS string  
solution of $D=5$ simple supergravity has a very similar structure to 
that of M5-brane. The metric of this solution is given by 
\begin{eqnarray}
ds_5^2 = H^{-1} (-dt^2 + dy^2) + H^2 (dr^2 + r^2 d\Omega_2),~~~
H\equiv 1+\frac Qr
\label{metric}
\end{eqnarray}
with the radial coordinate $r$ and the area element $d\Omega_2$ of the 
unit two-sphere. The $U(1)$ gauge field is
\begin{eqnarray} 
F^{ij}=-\sqrt{3} \epsilon^{ijk} H^{-4}\partial_k H
\label{U1} 
\end{eqnarray}
for the transverse space indices $i,j,k$, and $F^{MN}=0$ otherwise.

This string solution approaches Minkowski space as $r\rightarrow\infty$, 
and $AdS_3\times S^2$ as $r\rightarrow 0$.
The former statement is 
obvious. To see the latter, let us consider the new radial coordinate 
$r'$ such that 
\begin{eqnarray}
r'^2\equiv 4 Q^3 r^{-1}. 
\end{eqnarray} 
The horizon is located at $r'=\infty$. By using this $r'$, the metric 
can be written as
\begin{eqnarray}
ds_5^2 &=& \frac{4Q^2}{r'^2}
           \left( 1+ \frac{4Q^2}{r'^2} \right)^{-1} (-dt^2 + dy^2) 
          + \left( 1+ \frac{4Q^2}{r'^2} \right)^2
           \left(\frac{4Q^2}{r'^2}dr'^2 + Q^2 d\Omega_2 \right)\nonumber\\
&\stackrel{r'\rightarrow\infty}{\sim}&
 \frac{4Q^2}{r'^2}(-dt^2 + dy^2 +dr'^2) 
+ Q^2d\Omega_2.\label{metricAdS3xS2}
\end{eqnarray}
The first term of the right-hand side of (\ref{metricAdS3xS2}) is 
nothing but the $AdS_3$ metric in the holospheric coordinate.
We also see that the ratio of the radii of $AdS_3$ and $S^2$ is 
2:1. As we shall see in sect.3, this is the maximally 
supersymmetric configuration. Thus the string solution 
interpolates \cite{Gibbons:1993sv}
between two maximally supersymmetric vacua, just as M5-brane does, 
thereby justifying the name of ``solitonic string''.

\subsection{Energy and Spin in $AdS_3$}
The Anti-de Sitter space $AdS_3$ is the homogeneous space 
$SO(2,2)/SL(2,R)$. It can be thought of as the hyperboloid
\begin{eqnarray}
-u^2-v^2+x^2+y^2=-l^2
\end{eqnarray}
embedded in the flat space ${\bf R}^{4}$ with the metric 
\begin{eqnarray}
ds^2=l^2(-du^2-dv^2+dx^2+dy^2). \label{eq:metricmoto}
\end{eqnarray}
$l(>0)$ is the size of $AdS_3$.
Several convenient parameterizations are known. Among them we  
take
\begin{eqnarray}
&&u=l\cosh\rho \cos\tau,~~~v=l\cosh\rho \sin\tau,~~~\\
&&x=l\sinh\rho \cos\phi,~~~y=l\sinh\rho \sin\phi. 
\end{eqnarray}
Then the metric (\ref{eq:metricmoto}) becomes 
\begin{eqnarray}
ds^2=l^2(d\rho^2-\cosh^2\rho d\tau^2+\sinh^2\rho d\phi^2). 
\label{eq:AdSmetric}
\end{eqnarray}
It has the topology $S^1\times {\bf R}^{2}$
covered by 
$-\frac{\pi}2\leq\tau\leq\frac{3\pi}2$ and  
$\rho\geq 0$, $0\leq\phi\leq 2\pi$, respectively. 
One usually considers its universal covering space to unwrap 
the closed timelike curve along $\tau$. In this case $\tau$ runs over 
$-\infty<\tau<\infty$, and the topology becomes ${\bf R}^3$. 
Null and space-like infinity is represented by $\rho\rightarrow\infty$;
it has a topology of a cylinder.

It is well-known that $SO(2,2)$ is not a simple group, but a direct 
product $SL(2,{\bf R})\times SL(2,{\bf R})$. 
Let $M_{ij}$ $(i,j=0,1,2,3)$ be the generators of $SO(2,2)$ satisfying
\begin{eqnarray}
{[}M_{ij},~M_{kl}{]}=
\eta_{jk}M_{il}+\eta_{il}M_{jk}-\eta_{jl}M_{ik}-\eta_{ik}M_{jl}
\end{eqnarray}
with $\eta_{ij}={\rm diag}(-1,+1,+1,-1)$.
The indices $i,j,\ldots$ are raised by $\eta^{ij}$.
Defining 
\begin{eqnarray}
M^i_\pm\equiv\frac12\epsilon^{ijk}M_{\pm ij},~~~
M_\pm^{ij}\equiv M^{ij} \pm \frac12 \epsilon^{ijkl}M_{kl}~~~
(i,j,\ldots = 0,1,2),
\end{eqnarray}
each set of $M^i_\pm$ $(i=0,1,2)$ satisfies the commutation relations 
of $SO(2,1)\cong SL(2,{\bf R})$ Lie algebra.

In general, the energy and the spin of a particle in a $(d+1)$-dimensional 
Anti-de Sitter space are defined as the eigenvalue of $M^{0\;d+1}$ and 
the representation of the $SO(d)$ subgroup, respectively. 
Thus the energy $E$ and 
the spin $S$ in $AdS_3$ are labeled by two $SO(2)\cong U(1)$-charges 
$M^{03}$ and $M^{12}$. Therefore, one may identify
\begin{eqnarray}
L_0=-\frac i2 M_+^0,~~L_{\pm 1}=\frac i2 (M_+^2 \mp iM_+^1),\nonumber\\
\overline{L}_0=-\frac i2 M_+^0,
~~\overline{L}_{\pm 1}=\frac i2 (M_+^2 \mp iM_+^1)
\label{L0pm1}
\end{eqnarray}
with
\begin{eqnarray}
E=L_0 + \overline{L}_0,~~~S=L_0 - \overline{L}_0.
\end{eqnarray}
They induce the special conformal transformations on the boundary. 
\footnote{Note that $L_{\pm 1}$ and $\overline{L}_{\pm 1}$ 
are ${\it complex}$ linear 
combinations of $M_\pm^i$; they cannot be expressed with pure imaginary 
(nor real) coefficients alone. This means that each of two 
$SL(2,{\bf R})$ factors of $SO(2,2)$ is {\em not} the 
special conformal subgroup itself. Rather, one needs to go to its 
complexification $SL(2,{\bf C})$, and then consider another 
real form. These two real forms are related by an inner automorphism in 
this $SL(2,{\bf C})$. Of course, the origin of this ``twist'' 
is the fact that the boundary metric has a {\em Lorentzian} signature, 
whereas the Virasoro algebra is the symmetry of {\em Euclidean} conformal 
field theories.} 

%
%
\setcounter{equation}{0}
\section{Kaluza-Klein Mass Spectrum}

Let us now consider the compactification of $D=5$ simple supergravity 
on $S^2$. We will first derive 
a Freund-Rubin-like solution \cite{Freund:1980xh}.

The Lagrangian of $D\!=\!5$ simple supergravity is given by 
\begin{eqnarray}
{\cal L}&=&e_{5}\left[~-\frac14 R
~-\frac14 F_{MN}F^{MN} \right.\nonumber\\
&&-\frac i2\left(\overline{\psi}_M\widetilde{\Gamma}^{MNP}
         D_N\left(\frac{3\omega-\widehat{\omega}}2\right)\psi_P
         +\overline{\psi}_P\stackrel{\leftarrow}{D}_N
         \left(\frac{3\omega-\widehat{\omega}}2\right)
         \widetilde{\Gamma}^{MNP}\psi_M\right)\nonumber\\
&&-\frac1{6\sqrt{3}}e_5^{-1}\epsilon^{MNPQR}F_{MN}F_{PQ}A_R\nonumber\\
&&\left.-\frac{\sqrt{3}i}8\psi_M(\widetilde{\Gamma}^{MNPQ}
+2g^{M[P}g^{Q]N})\psi_N(F_{PQ}+\widehat{F}_{PQ})\right],
\end{eqnarray}
where we adopt the notation used in \cite{Chamseddine:1980sp}.
$\widetilde{\Gamma}^M$ are five-dimensional gamma matrices. 
In the background $\psi_M\!\!=\!0$, 
Einstein's and Maxwell's equations read, respectively,
\begin{eqnarray}
&& R_{MN}-\frac12 g_{MN}R = -(2F_{MP}F_N^{~~P}-\frac12 g_{MN}F^2),\\
&&F^{MN}_{~~~~;M}=\frac1{2\sqrt{3}}e_{5}^{-1}\epsilon^{NPQRS}F_{PQ}F_{RS},
\end{eqnarray}
where $;$ denotes the covariant derivative.
We write $\mu,\nu,\ldots =0,1,2$ as three - dimensional spacetime indices,  
and $m,n,\ldots=3,4$ as $S^2$ indices. 
The basic assumptions are  
\begin{eqnarray}
g_{\mu m}=F_{\mu m}=0,~~F_{mn}=fe_2\epsilon_{mn} 
\label{background}
\end{eqnarray}
for some non-zero real constant $f$. 
We also require that the Killing spinor equation 
\begin{eqnarray}
\delta\psi_M=0=\epsilon_{;M}
+\frac1{4\sqrt{3}}(\widetilde{\Gamma}_M^{~~PQ}
                   -4\delta_M^P\widetilde{\Gamma}^Q)F_{PQ}
                   \epsilon
\end{eqnarray}
be satisfied by maximally many supersymmetry parameters 
factorized in the form $\epsilon(x^\mu,y^m)=\epsilon(x^\mu)\eta(y^m)$
with three- and two-dimensional Dirac spinors $\epsilon(x^\mu)$ and 
$\eta(y^m)$, respectively. 
Taking the representation  
\begin{eqnarray}
\widetilde{\Gamma}^\mu=\gamma^\mu\otimes\Gamma^5,~~~
\widetilde{\Gamma}^m={\bf 1}\otimes\Gamma^m,~~~
\Gamma^5=i\Gamma^3\Gamma^4
\end{eqnarray}
in terms of three- and two-dimensional 
gamma matrices $\gamma^\mu$ and $\Gamma^m$, we obtain  
\begin{eqnarray}
&&\eta_{;m}-\frac i{\sqrt{3}} f \Gamma^5\Gamma_m\eta=0,\nonumber\\
&&\epsilon_{;\mu}-\frac i{2\sqrt{3}} f \gamma_\mu\epsilon=0,
\label{Killingspinor}\\
&&F_{\mu\nu}=0 \nonumber
\end{eqnarray}
in the background (\ref{background}). Then it follows from 
the the integrability conditions for (\ref{Killingspinor}) 
that the metric must satisfy: 
\begin{eqnarray}
R_{mnpq}&=&-\frac43 f^2(g_{mp}g_{nq}-g_{mq}g_{np}),\nonumber\\
R_{\mu\nu\rho\sigma}&=&+\frac13 f^2(g_{\mu\rho}g_{\nu\sigma}
                              -g_{\mu\sigma}g_{\nu\rho}).
\label{AdS3xS2}
\end{eqnarray}
Therefore, the background metric is that of $AdS_3\times S^2$.
We henceforth set $f={\sqrt{3}\over2}$ so that the radius of $S^2$ 
is normalized to be 1.

\subsection{Boson Masses}

The next step is the linearization of the field equations. Let us 
separate the fluctuations around the background as 
\begin{eqnarray}
g^{MN}&=&\gmaru^{MN}+h^{MN},\nonumber\\
F_{MN}&=&\Fmaru_{MN}+2a_{[N;M]},
\end{eqnarray}
where the fields with $\circ$ stand for the background 
(\ref{background})(\ref{AdS3xS2}), and the notations  
$\etamaru_{mn}=\emaru_{2}\epsilon_{mn}$ and 
$\etamaru_{\mu\nu\rho}=\emaru_{3}\epsilon_{\mu\nu\rho}$ 
will be used later. 
By writing out three- and two-dimensional indices, we find from
Einstein's equation:
\newcounter{tmp}
\setcounter{tmp}{\arabic{equation}}
\setcounter{equation}{0}
\def\theequation{E\arabic{equation}}

\begin{eqnarray}
&&-\frac12(h^M_{~~\mu;\nu;M}+h^M_{~~\nu;\mu;M}
         -h^M_{~~M;\mu;\nu}-h_{\mu\nu~~~;M}^{~~~;M})\nonumber\\
&&\hspace{2cm}=-\frac12 h_{\mu\nu}
 +\frac2{\sqrt{3}}\gmaru_{\mu\nu}\etamaru^{pq}a_{q;p},
\label{E1}
\\
&&-\frac12(h^M_{~~\mu;m;M}+h^M_{~~m;\mu;M}
         -h^M_{~~M;\mu;m}-h_{\mu m~~~;M}^{~~~;M})\nonumber\\
&&\hspace{2cm}=-2\sqrt{3}\etamaru_m^{~p}a_{[p;\mu]}-\frac12 h_{\mu m},
\label{E2}
\\
&&-\frac12(h^M_{~~m;n;M}+h^M_{~~n;m;M}
         -h^M_{~~M;m;n}-h_{mn~~~;M}^{~~~;M})\nonumber\\
&&\hspace{2cm}=-2\sqrt{3}(\etamaru_n^{~p}a_{[p;m]}+\etamaru_m^{~p}a_{[p;n]})
+\frac2{\sqrt{3}}\gmaru_{mn}\etamaru^{pq}a_{q;p}
+h_{mn},
\label{E3}
\end{eqnarray}
and from Maxwell's equations:
\setcounter{equation}{0}
\def\theequation{M\arabic{equation}}
\begin{eqnarray}
&&a^{[\mu;M]}_{~~~~~~;M}-\frac{\sqrt{3}}4 h^{\mu m;n}\etamaru_{mn}
-\etamaru^{\mu\nu\rho}a_{\rho;\nu}=0,
\label{M1}
\\
&&a^{[m;M]}_{~~~~~~;M}+\frac{\sqrt{3}}4 (-h^{mp;q}\etamaru_{pq}
+ h^{pM}_{~~~;M}\etamaru_{p}^{~m}
-\frac12 h^M_{~~M;p}\etamaru^{pm})=0.
\label{M2}
\end{eqnarray}
\setcounter{equation}{\thetmp}
\renewcommand{\theequation}{\arabic{section}.\arabic{equation}}

We fix the five-dimensional $U(1)$ gauge- and diffeomorphism 
degrees of freedom by imposing 
\begin{eqnarray}
a_m^{~~;m}=0,~~h^{\mu m}_{~~~;\mu}=h^{\mu m}_{~~~;m}=0.
\label{gauge}
\end{eqnarray} 
It is known, however, that the latter five conditions do not fix 
all the diffeomorphisms. Indeed, the coordinate 
transformation on the $S^2$: 
\begin{eqnarray}
\delta y^m \equiv \xi^m =  u\cdot Y^{;m}
\end{eqnarray}
for an arbitrary function $u(x^\mu)$ on $AdS_3$ 
and any particular spherical harmonics $Y(y^m)$ 
preserves the gauge conditions (\ref{gauge}) invariant if it is accompanied 
by the coordinate transformation on the $AdS_3$:
\begin{eqnarray}
\delta x^\mu \equiv \xi^\mu =  -u^{;\mu}\cdot Y.
\end{eqnarray}
Since this $\xi^M$ causes the change of the metric 
\begin{eqnarray}
\delta h^{mn} = 2u\cdot Y^{;m;n},\label{gaugemode}
\end{eqnarray}
one may additionally impose 
\begin{eqnarray}
h^m_{~~m}=0 \label{traceless}
\end{eqnarray}
using this degree of freedom. 

To obtain the mass spectrum, one expands the fields in terms of 
spherical harmonics. Since the rank of $SU(2)$ is one, the only  
independent harmonics on $S^2$ are the scalar harmonics; 
all the vector or tensor harmonics can be expressed by covariant 
derivatives of the scalar harmonics \cite{Regge:1957td}. Thus the fields 
are expanded as   
\begin{eqnarray}
h_{\mu\nu}&=&\sum_k H_{\mu\nu}^{(k)}Y_{(k)},\nonumber\\
h_{\mu m}&=&\sum_k (B_{1\mu}^{(k)}Y_{(k);m}
                 +B_{2\mu}^{(k)}\etamaru_{mn}Y_{(k)}^{~~;n}),\nonumber\\
h_{mn}&=&\sum_k (\phi_1^{(k)}Y_{(k);m;n}
                 +\phi_2^{(k)}\etamaru_{(m}^{~~~l}Y_{(k);n);l}
                 +\phi_3^{(k)}\gmaru_{mn}Y_{(k)}),\nonumber\\
a_\mu&=&\sum_k a^{(k)}_\mu Y_{(k)},\nonumber\\
a_m&=&\sum_k (b_1^{(k)}Y_{(k);m} 
              +b_2^{(k)}\etamaru_{mn}Y_{(k)}^{~~;n}),
\label{bosonexpansions}
\end{eqnarray}
where $Y_{(k)}$ are the spherical harmonic functions on $S^2$ satisfying 
\begin{eqnarray}
\Delta Y_{(k)} = k(k+1) Y_{(k)} 
\end{eqnarray}
for the Laplacian $\Delta$ on $S^2$. 
By substituting the gauge conditions (\ref{gauge})(\ref{traceless}), we 
obtain 
\begin{eqnarray}
h^{\mu m}_{~~~;m}=0 &\Rightarrow& B_{1\mu}^{(k)}=0,\nonumber\\
h^{\mu m}_{~~~;\mu}=0 &\Rightarrow& B_{2\mu}^{(k);\mu}=0,\nonumber\\
h^m_{~~m}=0 &\Rightarrow& k(k+1)\phi_1^{(k)}+2\phi_3^{(k)}=0,\nonumber\\
a_m^{~~;m}=0 &\Rightarrow& b_1^{(k)}=0.
\end{eqnarray}
Since eq.(\ref{AdS3xS2}) for $R_{mn}$ does not allow 
$\phi_2^{(k)}$ but a constant, one may ignore it. 
Thus every bosonic linearized field (\ref{bosonexpansions}) is expressed 
by a single mode function.  In the calculations below, 
$h_{\mu\nu}$, $h_{\mu m},\ldots$ will be regarded as modes of some 
particular harmonics. 

\vskip 2ex
\noindent
\underline{\it Scalars}
\vskip 2ex
Let us define 
\begin{eqnarray}
h\equiv h^\mu_{~\mu},~~\varphi\equiv h_{\mu\nu}^{~~~;\mu;\nu},
~\psi\equiv h_{mn}^{~~;m;n},~~da\equiv \etamaru^{pq}a_{q;p}.
\end{eqnarray}
The trace of the equation (\ref{E1}) yields 
\begin{eqnarray}
(\!\bigBox+ \frac12\Delta +\frac12)h - \varphi -2\sqrt{3} da =0.
\label{E1'}
\end{eqnarray}
$\bigBox$ is the three-dimensional d'Alembertian. 
Taking covariant derivatives $\mbox{(\ref{E2})}^{;\mu;m}$, 
we obtain
\begin{eqnarray}
\bigBox\psi -\bigBox\Delta h +\Delta\varphi -2\sqrt{3}\bigBox da =0.
\label{E2'}
\end{eqnarray}
The two-dimensional curl of (\ref{M2}) gives  
\begin{eqnarray}
(\bigBox + \Delta)da -\frac{\sqrt{3}}4\Delta h =0.
\label{M2'}
\end{eqnarray}
Only two of four fields $h$, $\varphi$, $\psi$ and $da$ are 
shown to be independent. 
Indeed, the trace of (\ref{E3}) gives the constraint
\begin{eqnarray}
\frac12\Delta h - \psi +\frac8{\sqrt{3}}da=0.
\label{E3'}
\end{eqnarray}
Also, using (\ref{E3'}), one can eliminate $\psi$ in (\ref{E2'}) to 
find 
\begin{eqnarray}
\bigBox\left(-\frac12\Delta h+\frac{2\sqrt{3}}3 da\right)
+\Delta\varphi=0. \label{eq:constraint0}
\end{eqnarray}
Eqs.(\ref{E1'}), (\ref{M2'}) and (\ref{eq:constraint0})  
yield another constraint 
\begin{eqnarray}
-\left(\frac12\Delta+\frac32\right)h-\varphi+\frac{10\sqrt{3}}3 da=0.
\label{constraint}
\end{eqnarray}
Using this constraint (\ref{constraint}), we can eliminate 
$\varphi$ in (\ref{E1'}): 
\begin{eqnarray}
(\bigBox + \Delta +2)h-\frac{16\sqrt{3}}3 da=0. 
\label{E1''}
\end{eqnarray}
One may read off the mass matrix $M_{\mbox{\scriptsize scalar}}$ 
from (\ref{M2'}) and (\ref{E1''}): 
\begin{eqnarray}
M^2_{\mbox{\scriptsize scalar}}=\left[
\begin{array}{cc}
\displaystyle\Delta+2 &\displaystyle -\frac{16\sqrt{3}}3\\
\displaystyle -\frac{\sqrt{3}}4 \Delta &\displaystyle \Delta
\end{array} 
\right].
\end{eqnarray}
Diagonalizing $M^2_{\mbox{\scriptsize scalar}}$, we obtain the 
two towers of eigenvalues
\begin{eqnarray}
-m_{\rm scalar}^{2}=\bigBox=-(k^2-k),~~~-(k^2+3k+2). \label{eq:scalarmass}
\end{eqnarray}

\vskip 2ex
\noindent
\underline{\it Vectors}
\vskip 2ex
We find from the two-dimensional curl of (\ref{E2}) that 
\begin{eqnarray}
(\bigBox+\Delta+\frac12){\cal H}_\mu +2\sqrt{3}\Delta a_\mu =0
\label{E2''}
\end{eqnarray}
with
\begin{eqnarray}
{\cal H}_\mu\equiv \etamaru^{pq}h_{\mu q;p}, 
\end{eqnarray}
while from (\ref{M1}) we get 
\begin{eqnarray}
(\bigBox+\Delta-\frac12)a_\mu-2\mbox{rot}a_\mu 
+\frac{\sqrt{3}}2 {\cal H}_\mu=0, 
\label{M1'}
\end{eqnarray}
where   
\begin{eqnarray}
\mbox{rot} a^\mu\equiv \etamaru^{\mu\nu\rho}a_{\rho;\nu}. 
\end{eqnarray}
With the help of the identity for vector fields:
\begin{eqnarray}
(\mbox{rot})^2 = -\bigBox +\frac12,
\end{eqnarray}
we find from the equations (\ref{E2''})(\ref{M1'}) that the operator 
rot has four different eigenvalues $\omega$ for each $SU(2)$-charge 
(labeled by $k$): 
\begin{eqnarray}
\omega = -k-2,~~ k-1,~~ -k,~~ k+1.   \label{eq:vectormass}
\end{eqnarray}
This means that all the massive vectors are ``self-dual''
in the sense that they satisfy the first-order differential equation:
\begin{eqnarray}
\mbox{rot}a^\mu=\omega a^\mu
\end{eqnarray}
for some constant $\omega$. Therefore, while ordinary vector 
fields have two 
massive states in three-dimensions, they have only {\em one} dynamical 
degree of freedom. Such ``self-dual'' bosons in odd dimensions were first 
recognized in the $S^4$ compactification of 
$D=11$ supergravity in \cite{Townsend:1984xs}, 
and played an important role in the 
construction of $D=7$ gauged supergravity.
\vskip 2ex
\noindent
\underline{\it Gravitons}
\vskip 2ex
The $R_{\mu\nu}$ equation (\ref{E1}) yields 

\begin{eqnarray}
(\bigBox+\Delta-\frac12)h_{(\mu\nu)}=0
\end{eqnarray}
for the traceless transverse part of $h_{\mu\nu}$. Thus we obtain 
a single tower for massive gravitons:
\begin{eqnarray}
-m_{\rm graviton}^{2}=\bigBox=-(k^2+k-\frac12). \label{eq:gravitonmass}
\end{eqnarray}

\subsection{Fermion Masses}
We next turn to the fermions.
Up to quadratic terms, the field equation for $\psi_M$ is 
\begin{eqnarray}
-i\widetilde{\Gamma}^{MNP}\psi_{P;N}
-\frac{\sqrt{3}}{4}i(\widetilde{\Gamma}^{MNPQ}
                     +2g^{M[P}g^{Q]N})\psi_NF_{PQ}=0.
\label{5dRSeq}
\end{eqnarray}
In the background (\ref{background}), this yields the following system  
of equations 
\begin{eqnarray}
&&\gamma^{\mu\nu\rho}\psi_{\rho;\nu}
+\gamma^{\mu\nu}\Gamma^5\Gamma^m(\psi_{m;\nu}-\psi_{\nu;m})
+\gamma^\mu\Gamma^{mn}\psi_{n;m}
-\frac34 i \gamma^{\mu\nu}\psi_\nu
=0,\\
&&\gamma^{\mu\nu}\Gamma^m\psi_{\nu;\mu}
-\Gamma^{mn}\Gamma^5\gamma^\nu(\psi_{n;\nu}-\psi_{\nu;n})
+\frac34 e_{2}^{-1}\epsilon^{mn}\psi_n
=0,
\end{eqnarray}
in which $\psi_\mu$ and $\psi_m$ are mixed. To decouple one from the 
other in the field equations, we take a convenient gauge-fixing 
condition for local supersymmetry \cite{Casher:1984ym,EN} 
\begin{eqnarray}
\widetilde{\Gamma}^M\psi_M=0. \label{gauge2}
\end{eqnarray}
Using the field equation (\ref{5dRSeq}), one finds 
\begin{eqnarray}
\psi^M_{~~;M}=-\frac14 i\gamma^\mu\psi_\mu
=\frac14 i\Gamma^5\Gamma^m\psi_m. 
\end{eqnarray}
This enables us to rewrite the system of field equations as 
\begin{eqnarray}
&&\gamma^\nu\psi^\mu_{~~;\nu}
-\frac i2 \gamma^{\mu\nu}\psi_\nu
+\frac i4\psi^\mu + \Gamma^5\Gamma^m\psi^\mu_{~~;m}=0,
\label{RSeq}\\
&&\gamma^\mu\psi^m_{~~;\mu}+\Gamma^5\Gamma^n\psi^m_{~~;n}
+i(\frac14\psi^m +\Gamma^{mn}\psi_n)=0,
\label{Diraceq}
\end{eqnarray}
where $\psi_\mu$ and $\psi_m$ appear in the separate equations.

In fact, the condition (\ref{gauge2}) again does not fix all the 
freedom of the 
supersymmetry. Using this residual degree 
of freedom, one may additionally set \cite{Casher:1984ym,EN}:
\begin{eqnarray} 
\Dmaru_m\psi^m\equiv\psi^m_{~~;m}-\frac i2\Gamma^5\Gamma_m\psi^m=0.
\label{spurious}
\end{eqnarray}
We will use this condition shortly.

\vskip 2ex
\noindent
\underline{\it Gravitini}
\vskip 2ex
The equation (\ref{RSeq}) determines the mass spectrum for the 
spin-$\frac32$ fields. 
Let us consider the eigenvalue problem for the Dirac operator 
on the two-sphere: 
\begin{eqnarray}
\Gamma^5\Gamma^mD_m\varphi = i\zeta\varphi,
\label{eigenvalueproblem}
\end{eqnarray}
where $D_m$ is the covariant derivative for a two-component complex spinor
$\varphi$ on the two-sphere. We assume the form of $\varphi$ as 
\begin{eqnarray}
\varphi(y) &=& a Y \eta(y)
               +b \partial_mY \Gamma^m\eta(y)\nonumber\\
            && +c Y \Gamma^5\eta(y)
               +d \partial_mY \Gamma^m\Gamma^5\eta(y),
\label{phi}
\end{eqnarray}
where $\eta(y)$ is the two-dimensional part of the Killing spinor 
(\ref{Killingspinor}). Plugging (\ref{phi}) into (\ref{eigenvalueproblem}),
one obtains the characteristic polynomial equation 
\begin{eqnarray}
\det\left[\begin{array}{cccc}
-i\zeta-i&0&0 & k(k+1)\\
0 & -i\zeta & -1 & 0 \\
0 & k(k+1) & -i\zeta+i & 0 \\
-1 & 0 & 0 & -i\zeta
\end{array}\right]
=0.
\end{eqnarray}
The solutions are 
\begin{eqnarray}
\zeta = \pm k,~~~ \pm (k+1). \label{eq:gravitinomass}
\end{eqnarray}
Each eigenvalue corresponds to a gravitino with mass 
$|\zeta+\frac14|$:
\begin{eqnarray}
\gamma^\nu\psi^\mu_{~~;\nu}-\frac i2\gamma^{\mu\nu}\psi_\nu
+i(\zeta+\frac 14)\psi^\mu=0.
\end{eqnarray}

\vskip 2ex
\noindent
\underline{\it Spinors}
\vskip 2ex
The masses of spin-$\frac12$ particles are determined by the
eigenvalues of the mass operator in (\ref{Diraceq}):
\begin{eqnarray}
\Gamma^5\Gamma^nD_n\psi^m+i\Gamma^{mn}\psi_n= i\kappa \psi^m.
\label{eigenvalueproblem2}
\end{eqnarray}
In general, one needs eight spherical harmonics to expand $\psi^m$:
\begin{eqnarray}
\psi^m&=&a_1 Y^{;m}\eta ~+a_2 Y^{;m}\Gamma^5\eta
\nonumber\\
&&+a_3 (\Dslash Y^{;m})\eta ~+a_4 (\Dslash Y^{;m})\Gamma^5\eta
\nonumber\\
&&  
+a_5 \etamaru^{mn} Y_{;n}\eta ~+a_6 \etamaru^{mn} Y_{;n}\Gamma^5\eta
\nonumber\\
&& ~+a_7 \etamaru^{mn} (\Dslash Y_{;n})\eta
         +a_8 \etamaru^{mn} (\Dslash Y_{;n})\Gamma^5\eta .   
\label{psi_m}
\end{eqnarray}
This expansion includes spurious modes; 
one can remove them by using the condition (\ref{spurious}).
After a little calculation, it turns out that the following relations 
among the expansion coefficients hold: 
\begin{equation}
a_1=a_3=0,~~a_2=a_8,~~[k(k+1)-1]a_4-a_6-ia_7=0.
\end{equation}
Thus the expansion of the physical wave function can be written as  
\begin{eqnarray}
\psi^m_{\rm phys}&=&a_2 (Y^{;m}\Gamma^5\eta+\etamaru^{mn} (\Dslash Y_{;n})\Gamma^5\eta) 
\nonumber\\
&&~+a_4((\Dslash Y^{;m})\Gamma^5\eta+(k(k+1)-1)\etamaru^{mn} Y_{;n}\Gamma^5\eta) 
\nonumber\\
&& ~~+a_5 \etamaru^{mn} Y_{;n}\eta 
\nonumber\\
&&  ~~~+a_7 (\etamaru^{mn} (\Dslash Y_{;n})\eta-i\etamaru^{mn} Y_{;n}\Gamma^5\eta) .
\end{eqnarray}
Thus the problem is reduced to 
the following $4\times 4$ matrix eigenvalue equation 
\begin{eqnarray}
\det\left[\begin{array}{cccc}
i\kappa & 0 & -k(k+1) & 0\\
0 & i\kappa & 0 & k(k+1) \\
1 & 0 & i\kappa+i & 0 \\
0 & -1 & 0 & i\kappa-i 
\end{array}\right]
=0.
\end{eqnarray}
The solutions are 
\begin{equation}
\kappa = \pm k,~ \pm (k+1).
\label{eq:spinormass}
\end{equation}
With these solutions our Dirac equation is given by 
\begin{eqnarray}
\gamma^\mu\psi^m_{~~;\mu}+i(\kappa+\frac 14)\psi^m=0.
\end{eqnarray}

The Kaluza-Klein mass spectrum obtained in this section is summarized 
in Table.1.
%
%
\setcounter{equation}{0}
\section{$SU(1,1|2)$ Lie Superalgebra}
In the previous section we obtained the Kaluza-Klein spectrum of $D=5$ 
simple supergravity on $AdS_3\times S^2$. We now study the 
representation theory of the Lie superalgebra $SU(1,1|2)$ into which 
each Kaluza-Klein mode is to fit.

The Lie superalgebra $SU(1,1|2)$ is defined by the following 
super-commutation relations denoted by $\mbox{[},\}$ among the fourteen 
generators $X_{\mu}$ $(\mu=1,$
$2,\cdots,14)$ \cite{Kac,Vlad}. 
\begin{eqnarray}
\mbox{[} X_{\mu}, X_{\nu} \} \CS{=}
X_{\mu}X_{\nu}-(-1)^{p(\mu)p(\nu)}X_{\nu}X_{\mu} \NN\\
\CS{=} i\,f_{\mu\nu\rho}X_{\rho}
\end{eqnarray}
for some structure constants $f_{\mu\nu\rho}$ (see Appendix A).
The {\it fermion number} $p(\mu)$ is 0 if $\mu\in\{ 1,2,\cdots,6 \}$, 
or 1 if $\mu\in\{ 7,8,\cdots,14\}$. Several remarks are in order. First, 
the naively defined $SL(2|2)$ (the complexification of $SU(1,1|2)$) 
as the algebra of the supertraceless $4\times 4$ matrices  
necessarily contains the obvious central element 
${\bf 1}_{4}={\rm diag}(1,1,1,1)$. Therefore, one considers the  
residue class 
\begin{eqnarray}
\mbox{Supertraceless $4\times 4$ matrices}/\{ {\bf 1}_{4}\}. 
\end{eqnarray}
By $SU(1,1|2)$ we mean this 
quotient algebra in this paper.  
Secondly, the difference between  $SU(2|2)$  
and $SU(1,1|2)$ 
should be explained. Both are the real forms of $SL(2|2)$ and 
contain the three-dimensional 
bosonic subalgebra generated by $X_{\mu}$ ($\mu=1,2,3$). 
For the former algebra, 
the structure constants $f_{\mu\nu\rho}$ are the same as those of 
$SU(2)$, {\it i.e.} $f_{\mu\nu\rho}=\epsilon_{\mu\nu\rho}$ 
(Levi-Civita tensor) while they are 
those of $SU(1,1)\cong SL(2,{\bf R})$, 
$f_{123} = -1, f_{231}=+1, f_{312}=+1$ for the latter. 
\subsection{$N=4$ Minimal Superconformal Algebra}
The Lie superalgebra $SU(1,1|2)$ is the finite subalgebra  
$\{ L_{0},L_{\pm 1},T^{i}_{0}, G^{i}_{\pm 1/2}, 
{\overline G}^{i}_{\pm 1/2} \}$  
the minimal $N=4$ superconformal algebra \cite{Ademollo:1976wv}
in the Neveu-Schwarz sector.
The explicit super-commutation relations are:
\begin{eqnarray}
&&\LS L_{m},L_{n}\RS = (m-n)L_{m+n}+
   \textstyle{1\over 2}km(m^{2}-1)\delta_{m+n,0},\NN\\
&&\{ G^{a}_{r},G^{b}_{s}\} = \{ {\overline G}^{a}_{r},{\overline G}^{b}_{s}\}=0,\NN\\
&&\{ G^{a}_{r},{\overline G}^{b}_{s}\}=
2\delta^{ab}L_{r+s}-2(r-s)\sigma^{i}_{ab}T^{i}_{r+s}
+\textstyle{1\over 2}k(4r^{2}-1)\delta_{r+s,0},\label{eq:N=4}\NN\\ 
&&\LS T^{i}_{m},T^{j}_{n}\RS =i\epsilon^{ijk}T^{k}_{m+n}
+\textstyle{1\over 2}km\delta_{m+n,0}\delta^{ij},\\
&&\LS T^{i}_{m},G^{a}_{r}]=-\sh\sigma^{i}_{ab}G^{b}_{m+r}, \quad 
\LS T^{i}_{m},{\overline G}^{a}_{r}]=-\sh\sigma^{i*}_{ab}{\overline G}^{b}_{m+r},
\NN\\
&&\LS L_m,G_r^a]=(\textstyle{1\over 2}m-r)G_{m+r}^a, \quad
\LS L_m,{\overline G}^{a}_{r}]=(\textstyle{1\over 2}m-r){\overline G}^{a}_{m+r},\NN\\
&&\LS L_m,T^{i}_{n}]=-nT^{i}_{m+n},\NN
\end{eqnarray}
where $\sigma^{i}$ is the Pauli spin matrix, $m$ and $n$ run over integers, 
$r$ and $s$ are half odd integers, $a,b=1$ or $2$, and 
$i$ is the $SU(2)$ index taking the value 1, 2 or 3.

A state $\ket{\phi}$ is said {\it chiral primary} if 
\BE
G^{2}_{-1/2}\ket{\phi}={\overline G^{1}_{-1/2}}\ket{\phi}=0, 
\label{eq:ChiralPrimary}
\EE
\BE
G^{a}_{n+1/2}\ket{\phi}={\overline G}^{a}_{n+1/2}\ket{\phi}=0, \qquad{\rm for}\quad 
n\ge 0,\quad a=1,2.
\EE
A chiral primary state $\ket{\phi}$ satisfies  
\BE
L_{0}\ket{\phi}=T_{0}^{3}\ket{\phi}. \label{eq:h=q}
\EE
\subsection{Oscillator Realizations}
To construct the representations of the $SU(1,1|2)$ algebra, 
we will consider the explicit realizations of the algebra  
in terms of super-oscillators \cite{Gunaydin:1986fe,Guenaydin}. 
\vskip 2ex
\noindent
\underline{\it The doubleton representation}
\vskip 2ex
The smallest representation, the   
{\it doubleton} representation, is given in terms of the 
two boson-fermion pairs ($a$,$\alpha$) and ($b$, $\beta$) as 
\begin{eqnarray}
&&L_0=\sh(a^\dagger a+bb^\dagger),\quad
L_1=ab,\quad
L_{-1}=a^\dagger b^\dagger,\nonumber\\
&&T_0^1=\sh(\alpha^{\dagger}\beta+\beta^{\dagger}\alpha),\quad
T_0^2=\textstyle{1\over 2i}
(\alpha^{\dagger}\beta-\beta^{\dagger}\alpha),\quad
T_0^3=\sh ({\alpha^{\dagger}\alpha-\beta^{\dagger}\beta}),
\\
&&G^1_{\frac12}=\sqrt{2}b\alpha,\quad 
G^2_{\frac12}=\sqrt{2}b\beta,\quad
G^1_{-\frac12}=\sqrt{2}a^\dagger\alpha,\quad
G^2_{-\frac12}=\sqrt{2}a^\dagger\beta,
\nonumber\\
&&\overline{G}^1_{\frac12}=\sqrt{2}a\alpha^{\dagger},\quad
\overline{G}^2_{\frac12}=\sqrt{2}a\beta^{\dagger},\quad
\overline{G}^1_{-\frac12}=\sqrt{2}b^\dagger\alpha^{\dagger},\quad
\overline{G}^2_{-\frac12}=\sqrt{2}b^\dagger\beta^{\dagger},\NN
\end{eqnarray}
provided that  
\begin{eqnarray}
\nu_{1}\equiv a^\dagger a-b^\dagger b
+\alpha^{\dagger}\alpha+\beta^{\dagger}\beta=1.\label{eq:nube1}
\end{eqnarray}
Note that $\nu_{1}$ commutes with all these generators.
One may achieve the quotient algebra with respect to the center $\nu_{1}$ 
by imposing the condition (\ref{eq:nube1}) on the space of states. 
The restricted Fock space consists of the states with $\nu_1$-charge $+1$.
$a^\dagger$,$\alpha^\dagger$ and $\beta^\dagger$ carry charge $+1$, 
while $b^\dagger$ does $-1$. 
Introducing the Fock vacuum $\ket{0}$ with the property 
\begin{eqnarray}
a\ket{0}=b\ket{0}=\alpha\ket{0}=\beta\ket{0}=0, 
\end{eqnarray}
one can obtain a four-dimensional basis (Fig.1(a))
\BE
\ket{\sh,\sh}=\alpha^{\dagger}\ket{0}, \quad
\ket{\sh,-\sh}=\beta^{\dagger}\ket{0}, \quad
\ket{1,0}^{(1)}=a^{\dagger}\ket{0},\quad
\ket{1,0}^{(2)}=b^{\dagger}\beta^{\dagger}\alpha^{\dagger}\ket{0}.
\EE
The states with the lower $L_{0}$ are fermionic in the doubleton representation. 
One of the two lower $L_{0}$ states is a chiral primary state 
($h=q=\sh$), and the other is an anti-chiral primary state ($h=-q=\sh$), 
which are mapped onto each other by $T_{0}^{i}$. 
\vskip 2ex
\noindent
\underline{\it The quarteton, massless and massive representations}
\vskip 2ex
The above realization can be easily extended to a higher-dimensional 
one. For this, we replace a single boson-fermion pair by two or more  
pairs as  
\begin{eqnarray}
a\longrightarrow (a_{1},a_{2},\cdots,a_{r})&,&
b\longrightarrow (b_{1},b_{2},\cdots,b_{r}),\NN\\
\alpha\longrightarrow (\alpha_{1},\alpha_{2},\cdots,\alpha_{r})&,&
\beta\longrightarrow (\beta_{1},\beta_{2},\cdots,\beta_{r}),
\end{eqnarray}
where $r$ is an integer ($r=$2,3,...). The products of the generators 
are correspondingly replaced as
\BE
ab\longrightarrow a\!\cdot\! b\equiv\sum_{i=1}^{r}a_{i}b_{i}, \mbox{~~etc.}
\EE
and the constraint $\nu_{1}=1$ by
\BE
\nu_{r}=a^\dagger\!\cdot\! a-b^\dagger\!\cdot\! b
+\alpha^{\dagger}\!\cdot\!\alpha+\beta^{\dagger}\!\cdot\!\beta\equiv r.
\EE

The basis of the representation can be constructed in a similar way.
For instance, let us consider the case $r=2$, which we call  
{\it quarteton} representation (Fig.1(b)).\footnote{We give a special 
name to this representation because, as we will show in the next section, 
it is this representation that plays a similar role to that of the 
doubleton in the $7+4$ compactification; the doubleton constructed above 
does not appear in the spectrum of our $AdS_3\times S^2$ compactification.} 
The three lowest-$L_{0}$ states are 
\BE
\ket{1,1}=\alpha_{1}^{\dagger}\alpha_{2}^{\dagger}\ket{0},\quad
\ket{1,0}=(\alpha_{1}^{\dagger}\beta_{2}^{\dagger}+
\beta_{1}^{\dagger}\alpha_{2}^{\dagger})\ket{0},\quad
\ket{1,-1}=\beta_{1}^{\dagger}\beta_{2}^{\dagger}\ket{0},
\EE
which are bosonic. 
The level-one descendants consist of 
two states, each of which is doubly 
degenerated. They are
\BEA
\ket{\sth,\sh}^{(1)}=(a_{1}^{\dagger}\alpha_{2}^{\dagger}-
a_{2}^{\dagger}\alpha_{1}^{\dagger})\ket{0}&,&
\ket{\sth,\sh}^{(2)}=
(b_{1}^{\dagger}\alpha_{1}^{\dagger}\alpha_{2}^{\dagger}\beta_{1}^{\dagger}+
b_{2}^{\dagger}\alpha_{1}^{\dagger}\alpha_{2}^{\dagger}\beta_{2}^{\dagger})
\ket{0},\NN\\
\ket{\sth,-\sh}^{(1)}=(b_{1}^{\dagger}\beta_{2}^{\dagger}-
b_{2}^{\dagger}\beta_{1}^{\dagger})\ket{0}&,&
\ket{\sth,-\sh}^{(2)}=
(a_{1}^{\dagger}\alpha_{1}^{\dagger}\beta_{1}^{\dagger}\beta_{2}^{\dagger}+
a_{2}^{\dagger}\alpha_{2}^{\dagger}\beta_{1}^{\dagger}\beta_{2}^{\dagger})
\ket{0}.
\EEA
The unique level-two descendant is
\BE
\ket{2,0}=(a_{1}^{\dagger}\alpha_{2}^{\dagger}-
a_{2}^{\dagger}\alpha_{1}^{\dagger})
(b_{1}^{\dagger}\beta_{1}^{\dagger}+b_{2}^{\dagger}\beta_{2}^{\dagger})
\ket{0}.
\EE
One may easily see that all the states above are antisymmetric with 
respect to the exchange of the indices $1$ and $2$. 

The states in the massless ($r=4$) and general massive 
($r=6,8,\ldots$) representations 
are shown in Fig.1(c). They contain the massless and massive 
representations of $SU(1,1)$ constructed in \cite{Gunaydin:1986fe},
respectively. One may also 
construct similar representations for odd $r$, but (like the doubleton) 
they do not appear in the spectrum, either.
%
%
\setcounter{equation}{0}
\section{Conformal Weights of the Boundary Operators}
We will now calculate the conformal dimension of a boundary field 
which couples to each Kaluza-Klein mode in the bulk.
The metric of $\AdS$ is given by (\ref{eq:AdSmetric}) with $l=2$: 
\BE
ds^{2}=4(-\cosh^{2}\!\rho\, d\tau^{2}+\sinh^{2}\!\rho\, d\phi^{2}+d\rho^{2}). 
\label{eq:metric}
\EE
We consider a $p$-form ${\cal C}$ on $\AdS$ with the spin $|S|$ 
and a boundary field 
${\cal O}$ interacting with ${\cal C}$ at the boundary $\partial\AdS$. 
The interaction between them is given by 
\BE
\int_{\partial\AdS}{\cal C}\wedge{\cal O}.
\EE
Suppose that the wave function behaves as ${\cal C}\sim e^{\lambda\rho}$ 
near the boundary $\rho=\infty$. Then the sum of the left and right 
conformal dimensions is given by \cite{Witten:1998qj} 
\BE
h_{L}+h_{R}=\lambda+2-p,\qquad |h_{L}-h_{R}|=|S|, \label{eq:Mh}
\EE
where $h_R$ (resp. $h_L$) is the eivenvalue of $L_0$ (resp. 
$\overline{L}_0$) (\ref{L0pm1}).

\vskip 2ex
\noindent
\underline{\it Scalars}
\vskip 2ex
We will first study the asymptotic behavior of the Klein-Gordon 
field $\varphi$ on $\AdS$ obeying 
\BE
(\bigBox-m_{\rm scalar}^{2})\varphi=0. \label{eq:KleinGordon}
\EE
Using the explicit metric (\ref{eq:metric}), one can rewrite this 
as 
\BE
\left[
{1\over 4}
\left(
\partial_{\rho}^{2}+2{\cosh 2\rho\over\sinh 2\rho}\partial_{\rho}
-{1\over\cosh^{2}\!\rho}\partial_{\tau}^{2}
+{1\over\sinh^{2}\!\rho}\partial_{\phi}^{2}
\right) -m_{\rm scalar}^{2}
\right] \varphi=0 .
\EE
If the asymptotic form at the boundary is $\varphi\sim e^{\lambda\rho}$,
$\lambda$ must satisfy
\BE
\lambda^{2}+2\lambda-4m_{\rm scalar}^{2}=0.
\EE
Since a scalar has equal left and right conformal dimensions 
$h_{L}=h_{R}$ 
($p=|S|=0$ in eq.(\ref{eq:Mh})), 
the equation we will solve is 
\BE
2h_{L}=\lambda+2.
\EE
>From (\ref{eq:scalarmass}), there are two distinct masses 
$m_{\rm scalar}^{2}=k^{2}-k$ and $k^{2}+3k+2$ for a given integer $k$. 
They have the degeneracy $2k+1$ coming from the spherical 
harmonics $Y_{(k)}$.  We mean by ``$SU(2)$-charge $q$'' this degeneracy, 
$i.e.$ $q=k$ in these cases.

The conformal dimensions $h_{L}$, $h_{R}$ and the $SU(2)$-charges 
for a given $k$ are summarized in Table 1(a).
The $SU(2)$ multiplet with $m_{\rm scalar}^{2}=k^{2}-k$ proves 
to contain a chiral primary 
state $h_{R}=q$, while the other does not.
\vskip 2ex
\noindent
\underline{\it (Anti-)Self-dual Vectors}
\vskip 2ex
Next we consider the problem for the (anti-)self-dual vector fields 
$\varphi_{\mu}$ satisfying
\BE
({\rm rot}\,\varphi)^{\mu}\equiv\etamaru^{\mu\nu\rho}\varphi_{\rho,\nu}
=\omega\varphi^{\mu} \label{eq:SDem}
\EE
for some constant $\omega$.
Plugging the $\AdS$ metric (\ref{eq:metric}) into this equation, 
we find 
\BEA
({\rm rot}\,\varphi)^{\rho}\CS{=}{1\over 2\sinh 2\rho}
\partial_{\ls\tau\rp}\varphi_{\lp\phi\rs}
=\omega\varphi^{\rho}={1\over 4}\omega\varphi_{\rho},\NN\\
({\rm rot}\,\varphi)^{\tau}\CS{=}{1\over 2\sinh 2\rho}
\partial_{\ls\phi\rp}\varphi_{\lp\rho\rs}
=\omega\varphi^{\tau}=-{\omega\over 4\cosh^{2}\!\rho}\varphi_{\tau},
\label{eq:rotSD}\\
({\rm rot}\,\varphi)^{\phi}\CS{=}{1\over 2\sinh 2\rho}
\partial_{\ls\rho\rp}\varphi_{\lp\tau\rs}
=\omega\varphi^{\phi}={\omega\over 4\sinh^{2}\!\rho}\varphi_{\phi}\NN
\EEA
Taking the limit $\rho\rightarrow\infty$ in (\ref{eq:rotSD}), 
we obtain the following reduced equation for $\varphi_{\mu}$ 
at the boundary:
\BE
\varphi_{\rho}=0,\quad \partial_{\rho}\varphi_{\phi}=
2\omega \varphi_{\tau},\quad
\partial_{\rho}\varphi_{\tau}=2\omega \varphi_{\phi}.
\EE
Thus the asymptotics are 
\BEA
\varphi_{+}\equiv\varphi_{\tau}+\varphi_{\phi}\sim e^{2\omega\rho}\qquad
&{\rm if}&\qquad \omega >0 ,\NN\\
\varphi_{-}\equiv\varphi_{\tau}-\varphi_{\phi}\sim e^{-2\omega\rho}\qquad
&{\rm if}&\qquad \omega <0.
\EEA
Substituting $\lambda=2|\omega|$ and $p=|S|=1$ in (\ref{eq:Mh}), 
we obtain
\BE
h_{L}+h_{R}=2|\omega|+1,\qquad |h_{L}-h_{R}|=1.
\EE 
In (\ref{eq:vectormass}), 
we have found four towers of excitations 
$\omega = k+1,\, -k,\, k-1,\, -k-2$ for 
a given integer $k$. They have $SU(2)$-charge $q=k$, 
and $h_{R}>h_{L}$ ($h_{R}<h_{L}$) if $\omega>0$ 
($\omega<0$). 
Only in the cases for $\omega=-k$ and $k-1$, 
the (anti-)self-dual vector fields are 
chiral primary (Table 1(b)). 
\vskip 2ex
\noindent
\underline{\it Gravitons}
\vskip 2ex
The equation of motion for the graviton $\varphi_{\mu\nu}$ is
\BE
(\bigBox-m_{\rm graviton}^{2})\varphi_{\mu\nu}=0. 
\label{eq:KleinGordonGraviton}
\EE 
Plugging the metric (\ref{eq:metric}) again, we see that 
\BE
\varphi_{\rho\rho},\varphi_{\rho\tau},\varphi_{\rho\phi}
\ll
\varphi_{\tau\tau},\varphi_{\phi\phi},\varphi_{\tau\phi}.
\EE
Thus we solve
\BE
\left[ 
{1\over 4}\left(
\partial_{\rho}^{2}-2\partial_{\rho}-2\right)-m_{\rm graviton}^{2}
\right]
\varphi_{\mu\nu}=0
\qquad (\mu,\nu)=(\tau,\tau),(\phi,\phi),(\tau,\phi).
\label{eq:RKGgraviton}
\EE
The conformal dimensions are  shown, due to 
$p=|S|=2$, to be 
\BE
h_{L}+h_{R}=\lambda=1+(4m_{\rm graviton}^{2}+3)^{1/2}, \qquad 
|h_{L}-h_{R}|=2.
\EE
Using (\ref{eq:gravitonmass}), we obtain the conformal dimensions 
shown in Table 1(c).  
Again, the states in the bottom row contain a chiral primary state, 
while the states in the first row do not.
\vskip 2ex
\noindent
\underline{\it Spinors}
\vskip 2ex
Consider the Dirac equation
\BE
(\gamma^{\mu}\nabla_{\mu}+m_{1/2})\psi\equiv 
\left[
\gamma^{\mu}\left(
\partial_{\mu}+\textstyle{1\over 4}\omega^{ij}_{\mu}\gamma_{ij} 
\right)+m_{1/2}
\right] \psi
=0. \label{eq:Dirac}
\EE 
The non-zero components of the spin connection are 
\BE
\omega^{12}_{\phi}=\cosh\rho,\qquad
\omega^{02}_{\tau}=\sinh\rho,
\EE
where the indeces $(0,1,2)$ are those of the local Lorentzian coordinate 
with the signature $(-,+,+)$.
We adopt the convention for the gamma matrices:
\BE
\gamma_{0}=i\sigma_{2},\quad \gamma_{1}=\sigma_{1},\quad
\gamma_{2}=\sigma_{3},
\EE 
where $\sigma_{1,2,3}$ are Pauli's spin matrices. 
In this representation one can rewrite the Dirac equation 
(\ref{eq:Dirac}) as 
\BEA
0&=&(\gamma^{\mu}\nabla_{\mu}+m_{1/2})\psi\NN\\
&=&\!\!\left[
{1\over 2\cosh\rho}\gamma^{0}\left(
\partial_{\tau}+\texthalf\sinh\rho\gamma_{02}
\right)
+{1\over 2\sinh\rho}\gamma^{1}\left(
\partial_{\phi}+\texthalf\cosh\rho\gamma_{12}
\right)
+{1\over 2}\gamma^{2}\partial_{\rho}+m_{1/2}
\right] \psi \NN\\
&\rightarrow&
\left[ {1\over 2}(\partial_{\rho}+1)\sigma_{3} +m_{1/2}\right] \psi,
\EEA
which implies 
\BE
\lambda = 2|m_{1/2}|-1. 
\EE
In the previous section we have found (eq.(\ref{eq:spinormass}))
\BE
m_{1/2} = \kappa-\textstyle{1\over 4},\qquad
SU(2)\mbox{-charge}=|\kappa|-\frac12,\qquad
\kappa=\pm k,\pm (k+1)
\EE
for each integer $k$.
Using these values with $p=0$, $|S|=1/2$  
in (\ref{eq:Mh}), the conformal dimensions are calculated 
as shown in Table 1(e),   
where the rule that 
$h_{R}>h_{L}$ ($h_{R}<h_{L}$) if $\kappa$ is positive (negative)
is used, similarly to the case of the (anti-)self-dual vectors. 
\vskip 2ex
\noindent
\underline{\it Gravitini}
\vskip 2ex
Finally, we consider the conformal dimensions of the gravitini. 
We consider the Dirac equation 
with fully covariantized derivative 
\BEA
&&(\gamma^{\nu}D_{\nu}+m)\psi^{\mu}=0,\NN\\
&&D_{\nu}\psi^{\mu}\equiv
\left(\partial_{\nu}+\omega^{ij}_{\nu}\gamma_{ij}\right)\psi^{\mu}
+\Gamma_{\nu\chi}^{\,\,\,\,\,\,\mu}\psi^{\chi}.
\label{eq:Dirac32}
\EEA
Substituting the metric (\ref{eq:metric}) and taking the limit 
$\rho\rightarrow\infty$, one can show that 
$\psi^{\tau}$, $\psi^{\phi}$ and $\upsilon\equiv e^{-\rho}\psi^{\rho}$ have 
the same scaling behavior. The Dirac equation is reduced to 
\BEA
&&\left[ 
{1\over 2}\sigma_{3}(\partial_{\rho}+2)+\left( m+{1\over 4}\right) 
\right] \psi^{\tau}-{1\over 2}\sigma_{3}\psi^{\phi}+
\sigma_{-}\upsilon=0,\NN\\
&&\left[ 
{1\over 2}\sigma_{3}(\partial_{\rho}+2)+\left( m+{1\over 4}\right) 
\right] \psi^{\phi}-{1\over 2}\sigma_{3}\psi^{\tau}+
\sigma_{-}\upsilon=0,
\label{eq:rDirac32}\\
&&\left[ 
{1\over 2}\sigma_{3}(\partial_{\rho}+2)+\left( m+{1\over 4}\right) 
\right] \upsilon
+{1\over 2}\sigma_{-}\psi^{\tau}
-{1\over 2}\sigma_{-}\psi^{\phi}
=0,\NN
\EEA
where $\sigma_{\pm}=\sigma_{1}\pm i\sigma_{2}$.
After a lengthy calculation, the asymptotic form is shown to be
\BE
\psi^{\tau},\psi^{\phi},\upsilon\sim e^{(\lambda-1)\rho},\qquad
\lambda-1 = |2\zeta+\texthalves|-1.
\EE
Using (\ref{eq:gravitinomass}) with $p=1$, $|S|=3/2$, we obtain
Table 1(d). 
\setcounter{equation}{0}
\section{AdS/CFT Correspondence}
\subsection{$SU(1,1|2)_{R}\times SL(2,{\bf R})_{L}$ 
Structure of the Spectrum}

We will now show how these infinite towers of particles fit into 
representations of $SU(1,1|2)_{R}\times SL(2,{\bf R})_{L}$, 
the finite dimensional subalgebras of the right $N=4$ superconformal 
and left Virasoro algebras. 
Since the right supersymmetry does not change the left 
conformal dimension $h_L$, an irreducible supermultiplet consists of 
fields with a common value of $h_L$. 
Thus we first assemble the following fields having equal $h_{L}=k$:
\begin{itemize}

\item A scalar with $h_{L}=h_{R}=q=k$ 
({\bf S${}_{\mbox{\scriptsize I}}$}),

\item A pair of spinors with $h_{L}=k$, $h_{R}=k+\sh$, and $q=k-\sh$  
({\bf Sp${}_{\mbox{\scriptsize I}}$},
{\bf Sp${}_{\mbox{\scriptsize II}}$}),

\item A self-dual vector with $h_{L}=k$, $h_{R}=k+1$, and $q=k-1$ 
({\bf V${}_{\!\mbox{\scriptsize I}}$}).              
  
\end{itemize}
They have precisely the correct quantum numbers to fit into the 
three-floor trapezoid diagram in Fig.2(a)!  One state in the diagram is 
mapped to another state on the same horizontal line by the action of 
the $SU(2)$ subalgebra $T_{0}^{i}$ in $SU(1,1|2)$. 
Borrowing the result on the asymptotic Virasoro algebra in $\AdS$ 
\cite{Brown:1986nw}, 
one concludes that the above set of three fields corresponds to 
a single chiral primary multiplet of $N=4$ superconformal algebra.  
In this case the chiral primary state corresponds to the scalar with 
the highest $U(1)$ charge in the $SU(2)$ multiplet.
Similarly, one may group   
\begin{itemize}

\item An anti-self-dual vector with $h_{L}=k+1$, $h_{R}=k$, and $q=k$ 
({\bf V${}_{\!\mbox{\scriptsize II}}$}),              

\item A pair of spinors with $h_{L}=k+1$, $h_{R}=k+\sh$, and $q=k-\sh$  
({\bf Sp${}_{\mbox{\scriptsize III}}$},
{\bf Sp${}_{\mbox{\scriptsize IV}}$}),

\item A scalar with $h_{L}=h_{R}=k+1$, and $q=k-1$ 
({\bf S${}_{\mbox{\scriptsize II}}$}).

\end{itemize}
These three fit to a trapezoid diagram as shown in 
Fig.2(b) and correspond to another chiral primary superconformal multiplet.
The (highest component of the) self-dual vector corresponds to 
the chiral primary field in this case.

We will now turn to the graviton multiplets. A massive graviton in 
three dimensions has two dynamical degrees of freedom, $i.e.$ the 
two ``helicity'' states $S=h_R-h_L=\pm 2$. Each of them is mapped 
to a different conformal field on the boundary. The trio of 
\begin{itemize}

\item A self-dual vector field with $h_{L}=k-1$, $h_{R}=k$, and $q=k$ 
({\bf V${}_{\!\mbox{\scriptsize III}}$}),              

\item A pair of gravitini with $h_{L}=k-1$, $h_{R}=k+\sh$, and $q=k-\sh$
({\bf Go${}_{\mbox{\scriptsize I}}$},
{\bf Go${}_{\mbox{\scriptsize II}}$}),

\item A graviton with $h_{L}=k-1$, $h_{R}=k+1$, and $q=k-1$ 
({\bf G${}_{\mbox{\scriptsize I}}$})

\end{itemize}
contains the $S=+2$ graviton and fits to the trapezoid in Fig.2(c), 
whereas the other trio of
\begin{itemize}

\item A graviton with $h_{L}=k+2$, $h_{R}=k$, and $q=k$ 
({\bf G${}_{\mbox{\scriptsize II}}$}),

\item A pair of gravitini with $h_{L}=k+2$, $h_{R}=k+\sh$, and $q=k-\sh$
({\bf Go${}_{\mbox{\scriptsize III}}$},
{\bf Go${}_{\mbox{\scriptsize IV}}$}),

\item An anti-self-dual vector with $h_{L}=k+2$, $h_{R}=k+1$, and $q=k-1$
({\bf V${}_{\!\mbox{\scriptsize IV}}$})              

\end{itemize}
has the $S=-2$ graviton and corresponds to another diagram 
shown in Fig.2(d). The chiral primary state in Fig.2(c) is a self-dual 
vector field, while the one in Fig.2(d) comes from a graviton.

We have thus shown that all the infinite towers of the Kaluza-Klein 
spectrum fall into {\rm four} infinite series of chiral primary 
(short) multiplets of $SU(1,1|2)_{R}\times SL(2,{\bf R})_{L}$.

\subsection{Classification of the Mass Spectrum}
Our results for the mass spectrum obtained in the previous sections  
are summarized in Table~\ref{spectrum}. Several remarks are in order:
\vskip 2.0ex
\noindent
i) All the multiplets appearing in the spectrum are short ($i.e.$ 
chiral primary) multiplets. This phenomenon of multiplet-shortening 
\cite{Freedman:1984na} is an essential property of
 the compactifications of $D=11$ supergravity; 
otherwise the highest-spin state in a long multiplet would exceed the 
limit for the allowed spins. In contrast, even the long multiplet 
of $SU(1,1|2)$ is ``short enough'' to be fit within the range of 
the allowed spins. Nevertheless,  
our multiplets are all 
short. These results, in turn, 
imply that the infinite series of graviton 
multiplets (the top row of the Table~{\ref{spectrum}}) do not exhaust 
all the multiplets in the spectrum; each column for a fixed ``excitation 
level $n$'' is decomposed into four irreducible representations of 
$SU(1,1|2)_R\times SL(2,{\bf R})_L$.

\vskip 1.0ex
\noindent
ii) The zero-modes on the solitonic string \cite{Mizoguchi:1998wv} 
appear as a {\em quarteton} in the first column ($n=0$). 
There appears no doubleton
in the table. It should be compared with the case of the $S^4$ 
compactification of $D=11$ supergravity, 
where the zero-modes on M5-brane correspond to 
a doubleton. 
In fact, the quarteton is a pure gauge mode just like singleton 
or doubleton in the higher-dimensional cases; for $k=1$, 
$Y_{(1);m;n}$ is proportional to $g_{mn}Y_{(1)}$ so that the scalar 
mode $\phi_3^{(1)}$ can be gauged away.

\vskip 1.0ex
\noindent
iii) The second column ($n=1$) has the same matter field content 
as that of the $T^2$ compactification of $D=5$ simple 
supergravity. After the experience in the $T^7/S^7$ or $T^4/S^4$ 
compactification, one may naturally expect the existence of an 
$SO(3)$ gauged supergravity in $D=3$ \cite{Achucarro:1986vz}. Indeed, the 
self-dual vector transforms as {\bf 3}, the correct representation 
to be the $SO(3)\cong SU(2)$ gauge field.

\vskip 1.0ex
\noindent
iv) Finally, we would like to mention the possibility that the symmetry 
of the mass spectrum might not be the direct product 
$SU(1,1|2)\times SL(2,{\bf R})$ but $OSp(2,2|2;-1)$. 
As we discuss in Appendix B, 
the bosonic generators of $OSp(2,2|2;\alpha)$ form the direct product 
$SL(2,{\bf R})\times SL(2,{\bf R})\times SU(2)$.  
If we take the limit $\alpha\rightarrow -1$, one of the two $SL(2,{\bf R})$  
groups decouples from the rest, and the $OSp(2,2|2;-1)$ 
decomposes into the semi-direct $SU(1,1|2)\times SL(2,{\bf R})$.
It would be more natural if the decoupled $SL(2,{\bf R})$  
will be ``supplied'' as $SL(2,{\bf R})_L$ 
from the larger 
(though semi-direct product) supergroup $OSp(2,2|2;-1)$. More specifically, 
in the notations in Appendix B, one identifies
\BE
{\overline L}_{0}=-Q_{z}^{1},\quad {\overline L}_{\pm 1}=Q_{x}^{1}\pm iQ_{y}^{1}.
\EE
The analogy from the 7+4 compactification supports this picture, 
since the symmetry group is $OSp(6,2|4)$ in that case. 
The minimal ($i.e.$ $SU(2)$) $N=4$ superconformal algebra does not 
contain the generator which distinguishes 
$G_{\pm {1\over 2}}^{1}$ and ${\overline G}_{\pm {1\over 2}}^{2}$ 
($G_{\pm {1\over 2}}^{2}$ and ${\overline G}_{\pm {1\over 2}}^{1}$). 
Therefore, although $Q^{1}_{i}$ act on the supercharges 
as a rotation among these supercharges,
all our result on the multiplet structure goes without 
any modification. 
Such a nontrivial $SL(2,{\bf R})_L$ action on the supercharges 
in the right sector leads to a two-dimensional 
conformal field theory of an unconventional type.
It would be interesting to see the physical implication of the 
free parameter $\alpha$ in $OSp(2,2|2;\alpha)$. We leave the study 
on this point to future research, but we mention that an interesting 
realization of the parameter $\alpha$ in $D=5$ black holes has been 
discussed in \cite{Townsend:1998ci}.
%
\section{Summary}
\setcounter{equation}{0}
We have studied the Kaluza-Klein spectrum of $D=5$ simple 
supergravity on $S^2$ with special interest in the relation to 
a two-dimensional $N=4$ superconformal field theory. 
A maximally supersymmetric Freund-Rubin-like background 
$AdS_3\times S^2$ was found, and turned out to be the geometry 
near the horizon of the solitonic string in $D=5$ simple supergravity.

The Kaluza-Klein spectrum was obtained by closely following the 
well-known techniques developed in the $S^4$/$S^7$ compactification 
of $D=11$ supergravity. We found a single tower of particles for 
gravitons, 4 for vectors, 2 for scalars and 4 for each of 
spin-3/2 and spin-1/2 fields. All the vector excitations are 
``(anti-)self-dual''
with having only half of what the ordinary massive 
vector has as its dynamical degrees of freedom (namely 1).

We next developed the representation theory for the Lie superalgebra 
$SU(1,1|2)$. The oscillator method was used. A special care was taken 
for the central element which arises in the naive definition of $SU(1,1|2)$. 
The quotient algebra $SU(1,1|2)$ was realized in a restricted Fock space. 
We constructed doubleton, quarteton, massless and massive representations 
by using one, two, four and $r$ ($=6,8,\ldots$) pair(s) of super-oscillators.
They are all short (chiral primary) representations.

We then calculated the conformal weight of the boundary operator 
by estimating the asymptotic behavior of the wave function for each 
Kaluza-Klein field. We finally showed that all the towers of particles 
were classified into four infinite series of chiral primary representations 
of $SU(1,1|2)\times SL(2,{\bf R})$ (direct product), or $OSp(2,2|2;-1) 
\cong SU(1,1|2)\times SL(2,{\bf R})$ (semi-direct product).

\section*{Acknowledgments}
The authors would like to acknowledge M. Scheunert for useful
remarks. One of them (S.M.) also thanks H. Nicolai for kind 
hospitality at the Albert-Einstein-Institute Potsdam, 
where a part of this work was done. 
%
\section*{Appendix}
\renewcommand{\theequation}{A.\arabic{equation}}
\setcounter{equation}{0}
\appendix
\section{$SU(1,1|2)$ and the Finite Subalgebra of $N=4$ 
Superconformal Algebra}
The Lie superalgebra $SL(2|2)$ can be defined using $4\times 4$ 
supertraceless matrices
\BE
X=\left[\begin{array}{cc|cc}
x_{11}&x_{12}&x_{13}&x_{14}\\
x_{21}&x_{22}&x_{23}&x_{24}\\
\hline
x_{31}&x_{32}&x_{33}&x_{34}\\
x_{41}&x_{42}&x_{43}&x_{44}
\end{array}\right]
\EE  
with
\BE
{\rm str}\, X\equiv x_{11}+x_{22}-x_{33}-x_{44}=0.
\EE
$x_{ij}$ ($i,j=1,...,4$) are complex numbers. 
By definition the identity matrix
\BE
{\bf 1}_{4}={\rm diag}(1,1,1,1)
\EE
is supertraceless. 
Thus the algebra of $4\times 4$ supertraceless matrices contains a 
center generated by ${\bf 1}_{4}$.  
$SL(2|2)$ is defined as the quotient algebra divided by this 
central element.

Now we consider the real form $SU(1,1|2)$ of $SL(2|2)$.  
We choose the following fourteen matrices as a basis: 
\BEA
&&L_{0}={1\over 2}\left[\begin{array}{cc|cc}
1&0& &\bigzerou\\
0&-1& & \\
\hline
\bigzerou& & &\bigzerou\\
&&&
\end{array}\right],\quad  
L_{1}=\left[\begin{array}{cc|cc}
0&0& &\bigzerou\\
i&0& & \\
\hline
\bigzerou& & &\bigzerou\\
&&&
\end{array}\right],\quad  
L_{-1}=\left[\begin{array}{cc|cc}
0&i& &\bigzerou\\
0&0& & \\
\hline
\bigzerou& & &\bigzerou\\
&&&
\end{array}\right],\NN\\
&&T_{0}^{3}={1\over 2}\left[\begin{array}{cc|cc}
\bigzerou& & &\bigzerou\\
& & & \\
\hline
\bigzerou& &1&0\\
 & &0&-1
\end{array}\right],\quad  
T_{0}^{1}={1\over 2}\left[\begin{array}{cc|cc}
\bigzerou& & &\bigzerou\\
& & & \\
\hline
\bigzerou& &0&1 \\
 & &1&0
\end{array}\right],\quad  
T_{0}^{2}={1\over 2}\left[\begin{array}{cc|cc}
\bigzerou& & &\bigzerou\\
& & & \\
\hline
\bigzerou& &0&-i \\
 & &i&0
\end{array}\right],\NN\\
&&G_{1\over 2}^{1}=\sqrt{2}\left[\begin{array}{cc|cc}
\bigzerou& &0&0\\
& &i&0 \\
\hline
\bigzerou& &&\bigzerou \\
&&&
\end{array}\right],\quad
G_{1\over 2}^{2}=\sqrt{2}\left[\begin{array}{cc|cc}
\bigzerou& &0&0\\
& &0&i \\
\hline
\bigzerou&&&\bigzerou \\
&&&
\end{array}\right],\label{eq:SL(2|2)}\\
&&G_{-{1\over 2}}^{1}=\sqrt{2}\left[\begin{array}{cc|cc}
\bigzerou& &1&0\\
& &0&0 \\
\hline
\bigzerou& &&\bigzerou \\
&&&
\end{array}\right],\quad
G_{-{1\over 2}}^{2}=\sqrt{2}\left[\begin{array}{cc|cc}
\bigzerou& &0&1\\
& &0&0 \\
\hline
\bigzerou&&&\bigzerou \\
&&&
\end{array}\right],\NN\\
&&{\overline G}_{1\over 2}^{1}=\sqrt{2}\left[\begin{array}{cc|cc}
\bigzerou& &&\bigzerou\\
&&&\\
\hline
1&0&&\bigzerou \\
0&0&&
\end{array}\right],\quad
{\overline G}_{1\over 2}^{2}=\sqrt{2}\left[\begin{array}{cc|cc}
\bigzerou& &&\bigzerou\\
&&&\\
\hline
0&0&&\bigzerou\\
1&0&&
\end{array}\right],\NN\\
&&{\overline G}_{-{1\over 2}}^{1}=\sqrt{2}\left[\begin{array}{cc|cc}
\bigzerou& &&\bigzerou\\
&&&\\
\hline
0&i&&\bigzerou \\
0&0&&
\end{array}\right],\quad
{\overline G}_{-{1\over 2}}^{2}=\sqrt{2}\left[\begin{array}{cc|cc}
\bigzerou& &&\bigzerou\\
&&&\\
\hline
0&0&&\bigzerou\\
0&i&&
\end{array}\right],\NN
\EEA
It is straightforward to check that they satisfy (\ref{eq:N=4}). 
For example, 
\BEA
\{ G_{1\over 2}^{1},{\overline G}_{-{1\over 2}}^{1}\} &=&
\left[\begin{array}{cc|cc}
0&&&\\
&-2&&\\
\hline
&&-2&\\
&&&0
\end{array}\right]\NN\\
&\cong&
\left[\begin{array}{cc|cc}
1&&&\\
&-1&&\\
\hline
&&-1&\\
&&&1
\end{array}\right]
=2(L_{0}-T_{0}^3).
\EEA
The last line is an equality in the quotient algebra.
\renewcommand{\theequation}{B.\arabic{equation}}
\setcounter{equation}{0}
\section{$OSp(4|2)$, $OSp(4|2;\alpha)$ and $N=4$ Superconformal Algebras}
The complex Lie superalgebra $OSp(2m|2n)$ is defined as 
the superalgebra of $(2m+2n)\times (2m+2n)$ matrices $X$ in the form
\BE
X=\left( \begin{array}{c|c}
A&B\\
\hline
C&D
\end{array}\right),
\EE
where the $2m\times 2m$ and $2n\times 2n$ matrices $A$ and $D$
satisfy 
\BEA
&&^{t}DG+GD=0,\quad ^{t}A=-A,\quad B=-^{t}\!CG,\NN\\
&&G=\left[ \begin{array}{cc}
O&{\bf 1}_{n}\\
-{\bf 1}_{n}&O
\end{array}\right].
\EEA
${\bf 1}_{n}$ is the $n\times n$ identity matrix.
If and only if $m=2$ and $n=1$, the above defined $OSp(2m|2n)$ can be extended 
to a one-parameter family of Lie superalgebras $OSp(4|2;\alpha)$ 
with a real parameter $\alpha$. 
According to the notations in \cite{Vlad}, we give the definition of 
Lie superalgebra $OSp(4|2;\alpha)$. That is, among 
the seventeen generators $Q_{j}^{m}\,(j=x,y,z;m=1,2,3)$ and 
$R_{\mu\nu\rho}\,(\mu,\nu,\rho=1,2)$, the super-commutation 
relations are:
\BEA
&&[ Q_{j}^{m},Q_{k}^{n}] = i\delta_{mn}\epsilon_{jkl}Q_{l}^{m},\NN\\
&&[ Q_{j}^{1},R_{\mu\nu\rho}] = 
  \textstyle{1\over 2}\sigma_{\mu'\mu}^{j}R_{\mu'\nu\rho},\, 
[ Q_{j}^{2},R_{\mu\nu\rho}] = 
  \textstyle{1\over 2}\sigma_{\nu'\nu}^{j}R_{\mu\nu'\rho},\, 
[ Q_{j}^{3},R_{\mu\nu\rho}] = 
  \textstyle{1\over 2}\sigma_{\rho'\rho}^{j}R_{\mu\nu\rho'}, \NN\\
&&\{ R_{\mu\nu\rho},R_{\mu'\nu'\rho'} \} =
\alpha_{1}C_{\rho\rho'}C_{\nu\nu'}(C\sigma^{j})_{\mu\mu'}Q_{j}^{1}\NN\\
&&\hspace{3.5cm}+\alpha_{2}C_{\rho\rho'}(C\sigma^{j})_{\nu\nu'}
  C_{\mu\mu'}Q_{j}^{2}
+\alpha_{3}(C\sigma^{j})_{\rho\rho'}C_{\nu\nu'}C_{\mu\mu'}Q_{j}^{3} 
\EEA
with
\BE
\alpha_{1}+\alpha_{2}+\alpha_{3}=0
\EE
and
\BE
C=\left( \begin{array}{cc}
0&1\\ -1&0
\end{array}\right).
\EE
The nine bosonic elements $Q_{j}^{m}$   
consist of three mutually commuting $SL(2,{\bf C})$ algebras 
$SL(2,{\bf C})\times SL(2,{\bf C})\times SL(2,{\bf C})$, 
and the eight fermionic elements $R_{\mu\nu\rho}$ form 
a fundamental representation. 
Since the algebra depends on $\alpha_i$ only through the ratio 
\BE
\alpha_{2}=1.
\EE
Thus there is only one free parameter $\alpha=\alpha_3=-\alpha_1-1$ 
\cite{Kac}. 
When $\alpha=-1/2$, the algebra is reduced to the ordinary $OSp(4|2)$ 
algebra \cite{Kac,Vlad}.

The most important case for this paper is the $\alpha\rightarrow -1$ limit. 
Let us consider the real form $OSp(2,2|2;-1)$. In this case 
$Q_{j}^{m}$ form 
$SL(2,{\bf R})\times SL(2,{\bf R})\times SU(2)$. 
At $\alpha=-1$, $\alpha_i$ can be fixed as
\BE
\alpha_{1}=0,\quad\alpha_{2}=1,\quad\alpha_{3}=-1.
\EE
Let 
\BEA
&&T_{0}^{1}=Q_{x}^{2},\quad T_{0}^{2}=Q_{y}^{2},\quad T_{0}^{3}=Q_{z}^{2},\NN\\
&&L_{0}=-Q_{z}^{3},\quad L_{\pm 1}=Q_{x}^{3}\pm iQ_{y}^{3},\NN\\
&&G_{1\over 2}^{1}=R_{121},\quad G_{1\over 2}^{2}=-R_{111},\quad 
G_{-{1\over 2}}^{1}=R_{122},\quad G_{-{1\over 2}}^{2}=-R_{112},
\label{eq:a=-1/2}\\
&&{\overline G}_{1\over 2}^{1}=R_{211},\quad 
{\overline G}_{1\over 2}^{2}=-R_{221},\quad 
{\overline G}_{-{1\over 2}}^{1}=R_{212},\quad 
{\overline G}_{-{1\over 2}}^{2}=-R_{222},\NN
\EEA
then it can be shown that the $SL(2,{\bf R})$ generated by $Q_{1}^{i}$ 
decouples and that the other generators $\{ L_{0,\pm 1},T_{0}^{i},
G_{\pm 1/2}^{a}, {\overline G}_{\pm 1/2}^{a}\} $ form the closed algebra 
$SU(1,1|2)$ \cite{Vlad,VanDerJeugt:1985hq}. 
Note that the $SL(2,{\bf R})$ generated 
by $Q_{1}^{i}$ acts on $G$, $\overline{G}$ nontrivially, 
and hence the whole $OSp(2,2|2;-1)$ is a 
semi-direct product.

%

\newpage
\pagestyle{empty}
%
%
\begin{table}[p]
\caption{Conformal weights of the boundary fields corresponding 
to the Kaluza-Klein modes.}
\begin{center}
(a) Scalars.\\
\begin{tabular}{|c|c|ccc|}
\hline
Series &$m_{\rm scalar}^{2}$ & $h_{L}$ & $h_{R}$ & $q$ \\
\hline
{\bf S${}_{\mbox{\scriptsize I}}$}&$k^{2}-k$    & $k$   & $k$   & $k$ \\
{\bf S${}_{\mbox{\scriptsize II}}$}&$k^{2}+3k+2$ & $k+2$ & $k+2$ & $k$ \\
\hline
\end{tabular}
\vskip 2em
(b) Vectors.\\
\begin{tabular}{|c|c|ccc|}
\hline
Series&$\omega$ & $h_{L}$ & $h_{R}$ & $q$ \\
\hline
{\bf V${}_{\!\mbox{\scriptsize I}}$}&$k+1$  & $k+1$ & $k+2$ & $k$ \\
{\bf V${}_{\!\mbox{\scriptsize II}}$}&$-k$   & $k+1$   & $k$ & $k$ \\
{\bf V${}_{\!\mbox{\scriptsize III}}$}&$k-1$  & $k-1$   & $k$ & $k$ \\
{\bf V${}_{\!\mbox{\scriptsize IV}}$}&$-k-2$ & $k+3$ & $k+2$ & $k$ \\
\hline
\end{tabular}
\vskip 2em
(c) Gravitons.\\
\begin{tabular}{|c|c|ccc|}
\hline
Series&$m_{\rm graviton}^{2}$ & $h_{L}$ & $h_{R}$ & $q$ \\
\hline
{\bf G${}_{\mbox{\scriptsize I}}$}
 &$k^{2}+k-\textstyle{1\over 2}$  & $k$   & $k+2$ & $k$ \\
{\bf G${}_{\mbox{\scriptsize II}}$}
 &$k^{2}+k-\textstyle{1\over 2}$  & $k+2$ & $k$   & $k$ \\
\hline
\end{tabular}
\vskip 2em
(d) Gravitini.\\
\begin{tabular}{|c|c|ccc|c|}
\hline
Series&$\zeta$ & $h_{L}$ & $h_{R}$ & $q$\\
\hline
{\bf Go${}_{\mbox{\scriptsize I}}$}
 &$k$      & $k-1$ & $k+\texthalf$   & $k-\texthalf$\\
{\bf Go${}_{\mbox{\scriptsize II}}$}
 &$k+1$    & $k$   & $k+\texthalves$ & $k+\texthalf$\\
{\bf Go${}_{\mbox{\scriptsize III}}$}
 &$-k$     & $k+2$ & $k+\texthalf$   & $k-\texthalf$\\
{\bf Go${}_{\mbox{\scriptsize IV}}$}
 &$-(k+1)$ & $k+3$ & $k+\texthalves$ & $k+\texthalf$\\
\hline
\end{tabular}
\vskip 2em
(e) Spinors.\\
\label{tab:spinor}
\begin{tabular}{|c|c|ccc|}
\hline
Series&$\kappa$ & $h_{L}$ & $h_{R}$ & $q$\\
\hline
{\bf Sp${}_{\mbox{\scriptsize I}}$}
 &$k$      & $k$   & $k+\texthalf$   & $k-\texthalf$\\
{\bf Sp${}_{\mbox{\scriptsize II}}$}
 &$k+1$    & $k+1$ & $k+\texthalves$ & $k+\texthalf$\\
{\bf Sp${}_{\mbox{\scriptsize III}}$}
 &$-k$     & $k+1$   & $k+\texthalf$   & $k-\texthalf$\\
{\bf Sp${}_{\mbox{\scriptsize IV}}$}
 &$-(k+1)$ & $k+2$ & $k+\texthalves$ & $k+\texthalf$\\
\hline
\end{tabular}
\end{center}
\end{table}
\newpage
\renewcommand{\thetable}{\arabic{table}}
\setcounter{table}{1}
\begin{table}
\caption{Kaluza-Klein spectrum of $AdS_3\times S^2$ supergravity.}
\label{spectrum}
\begin{center}
\begin{tabular}{|c|c|c|ccccc|}
\hline
Field & Massive&$\AdS$ spin&\multicolumn{5}{|c|}{Multiplicities} \\
      & states($n\geq 2$)&$S=h_R-h_L$&$n$=0,& 1,& 2, & 3, &$\ldots$ \\
\hline
Graviton
({\bf G${}_{\mbox{\scriptsize I}}$})
        &1&$+2$&---& 1 & 3 & 5 & $\cdots$\\
Gravitino
({\bf Go${}_{\mbox{\scriptsize I}}$},
{\bf ${}_{\mbox{\scriptsize II}}$})
       &2&$+\frac32$&---& 2 & 4 & 6 & $\cdots$\\
Self-dual vector
({\bf V${}_{\!\mbox{\scriptsize III}}$})
&1&$+1$&---& 3 & 5 & 7 & $\cdots$\\
\hline
Graviton
({\bf G${}_{\mbox{\scriptsize II}}$})
        &1&$-2$&---&--- & 3 & 5 & $\cdots$\\  
Gravitino
({\bf Go${}_{\mbox{\scriptsize III}}$},
{\bf ${}_{\mbox{\scriptsize IV}}$})
       &2&$-\frac32$&---&---& 2 & 4 & $\cdots$\\
Anti-self-dual vector
({\bf V${}_{\!\mbox{\scriptsize IV}}$})
&1&$-1$  &---&---& 1 & 3 & $\cdots$\\
\hline  
Self-dual vector
({\bf V${}_{\!\mbox{\scriptsize I}}$})
&1&$+1$& 1 & 3 & 5 & 7 & $\cdots$\\
Spinor
({\bf Sp${}_{\mbox{\scriptsize I}}$},
{\bf ${}_{\mbox{\scriptsize II}}$})
          &2&$+\frac12$& 2 & 4 & 6 & 8 & $\cdots$\\
Scalar
({\bf S${}_{\mbox{\scriptsize I}}$})
          &1&$0$& 3 & 5 & 7 & 9 & $\cdots$\\
\hline
Anti-self-dual vector
({\bf V${}_{\!\mbox{\scriptsize II}}$})
&1&$-1$&---&---& 3 & 5 & $\cdots$\\
Spinor
({\bf Sp${}_{\mbox{\scriptsize III}}$},
{\bf ${}_{\mbox{\scriptsize IV}}$})
          &2&$-\frac12$&---&---& 2 & 4 & $\cdots$\\
Scalar
({\bf S${}_{\mbox{\scriptsize II}}$})
          &1&$0$&---&---& 1 & 3 &$\cdots$\\
\hline
\end{tabular}
\end{center}
\end{table}
%
%
\newpage
\section*{Figure Captions}
\noindent
Fig.1~ Representations of $SU(1,1|2)$.
(a) The doubleton representation.
(b) The ``quarteton'' representation ($r=2$).
(c) The quarteton ($r=2$), massless ($r=4$) 
and massive ($r=6,8,\ldots$) representations.
\vskip 2ex \noindent
Fig.2~ The boundary fields grouped into 
$SU(1,1|2)\times SL(2,{\bf R})$ 
(or $OSp(2,2|2;\alpha\!\rightarrow\! -1)$)
multiplets.
As $N=4$ superconformal fields, the chiral primary fields 
corresponds to 
(a) a scalar (b) an anti-self-dual vector (c) a self-dual vector 
(d) a graviton. 
%
%
\newpage
\begin{figure}
  \epsfxsize = 12 cm   
  \centerline{\epsfbox{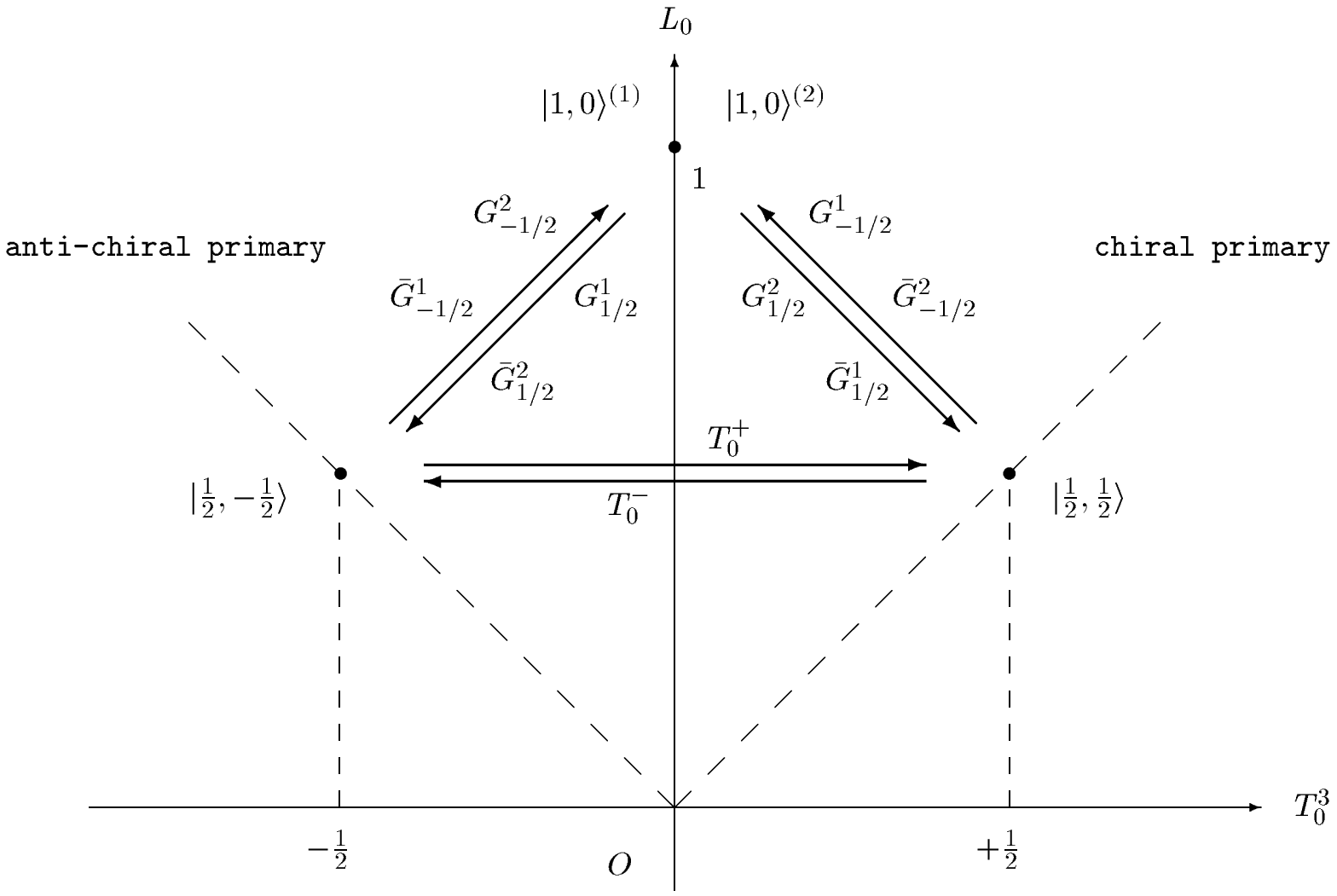}}
  \centerline{Fig.1(a)}
\end{figure}
\begin{figure}
  \epsfxsize = 12 cm   
  \centerline{\epsfbox{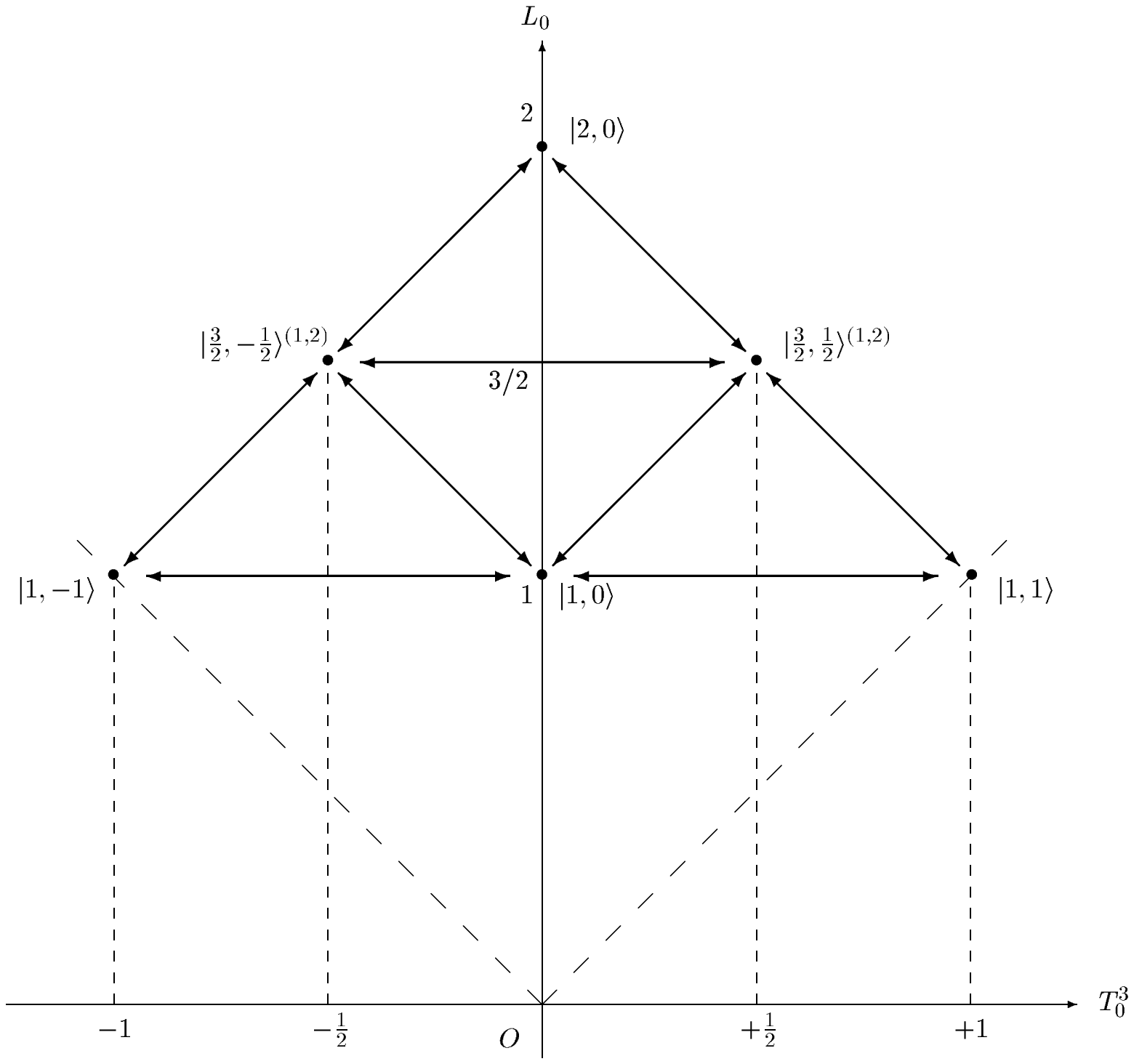}}
  \centerline{Fig.1(b)}
\end{figure}
\begin{figure}
  \epsfxsize = 16 cm   
  \centerline{\epsfbox{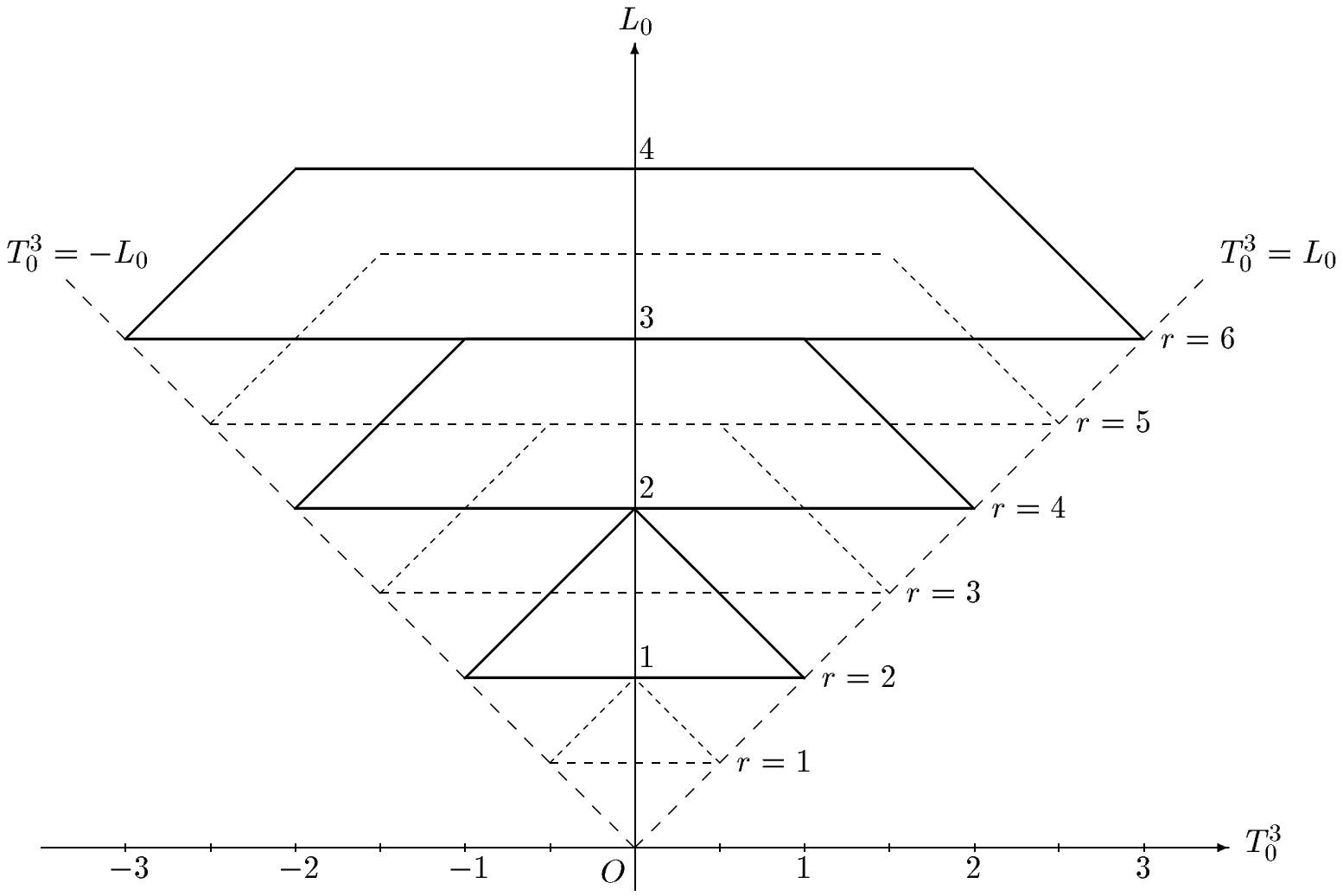}}
  \centerline{Fig.1(c)}
\end{figure}
~\
\newpage
\begin{figure}
  \epsfxsize = 12 cm   
  \centerline{\epsfbox{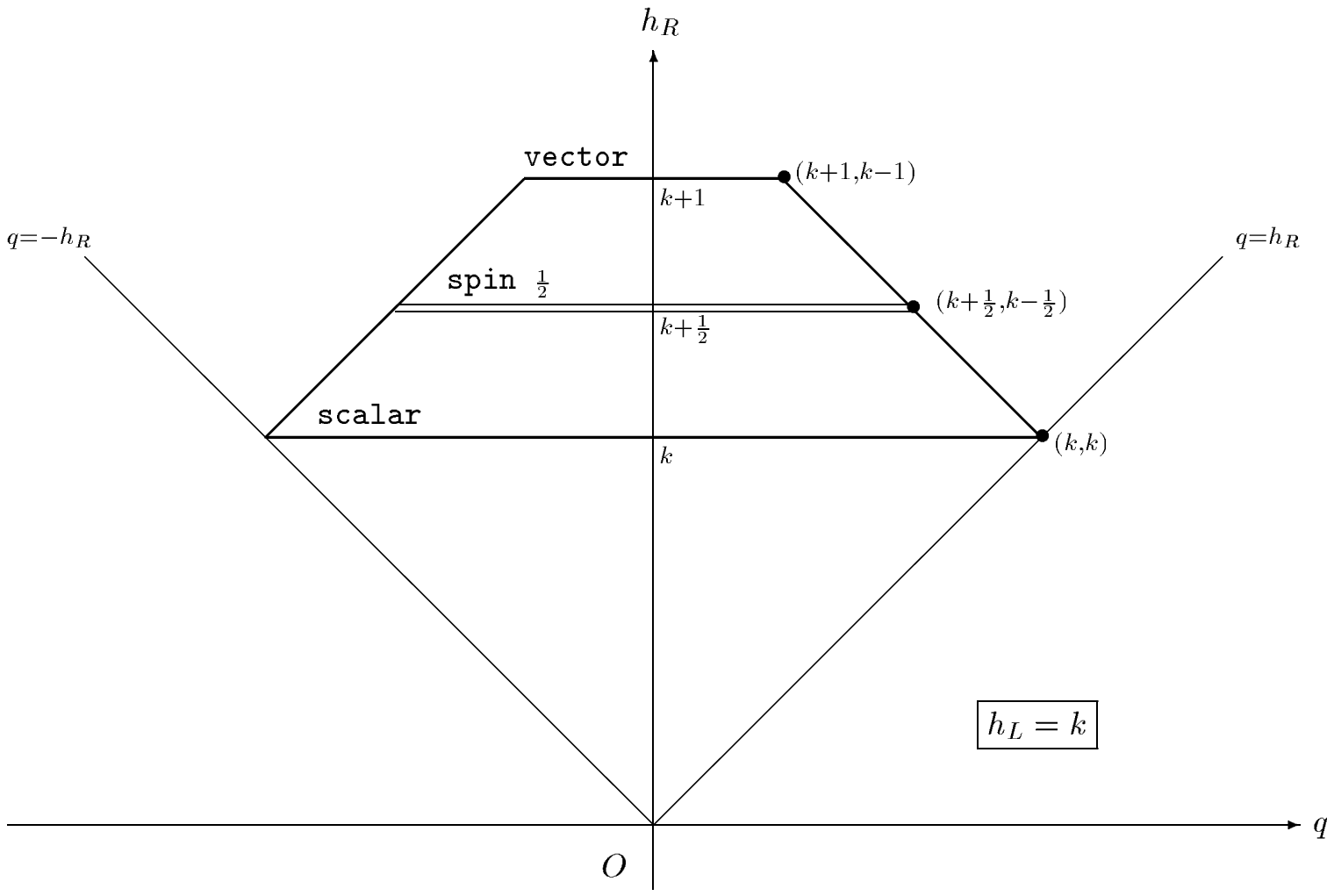}}
  \centerline{Fig.2(a)}
\end{figure}
\begin{figure}
  \epsfxsize = 12 cm   
  \centerline{\epsfbox{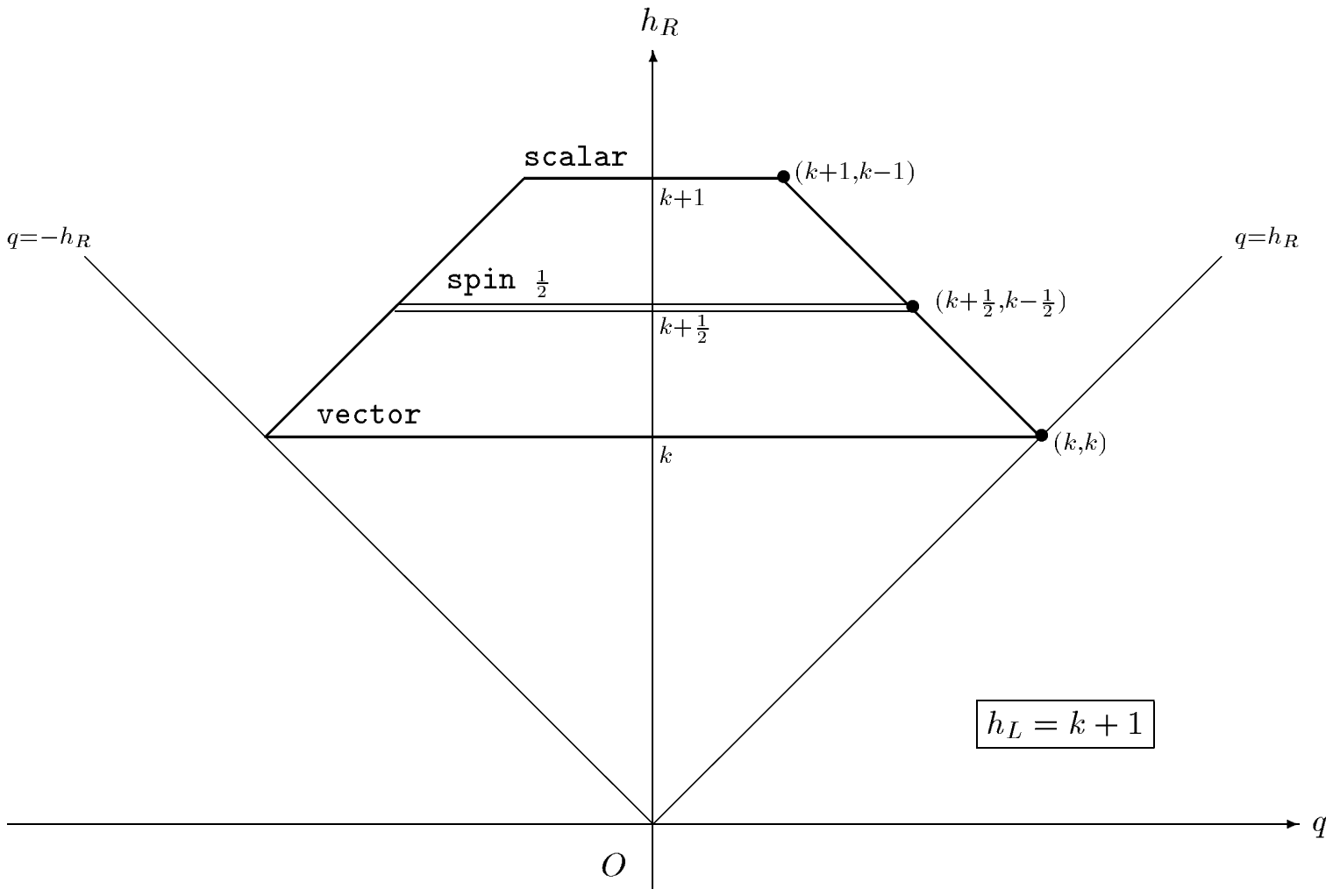}}
  \centerline{Fig.2(b)}
\end{figure}
\begin{figure}
  \epsfxsize = 12 cm   
  \centerline{\epsfbox{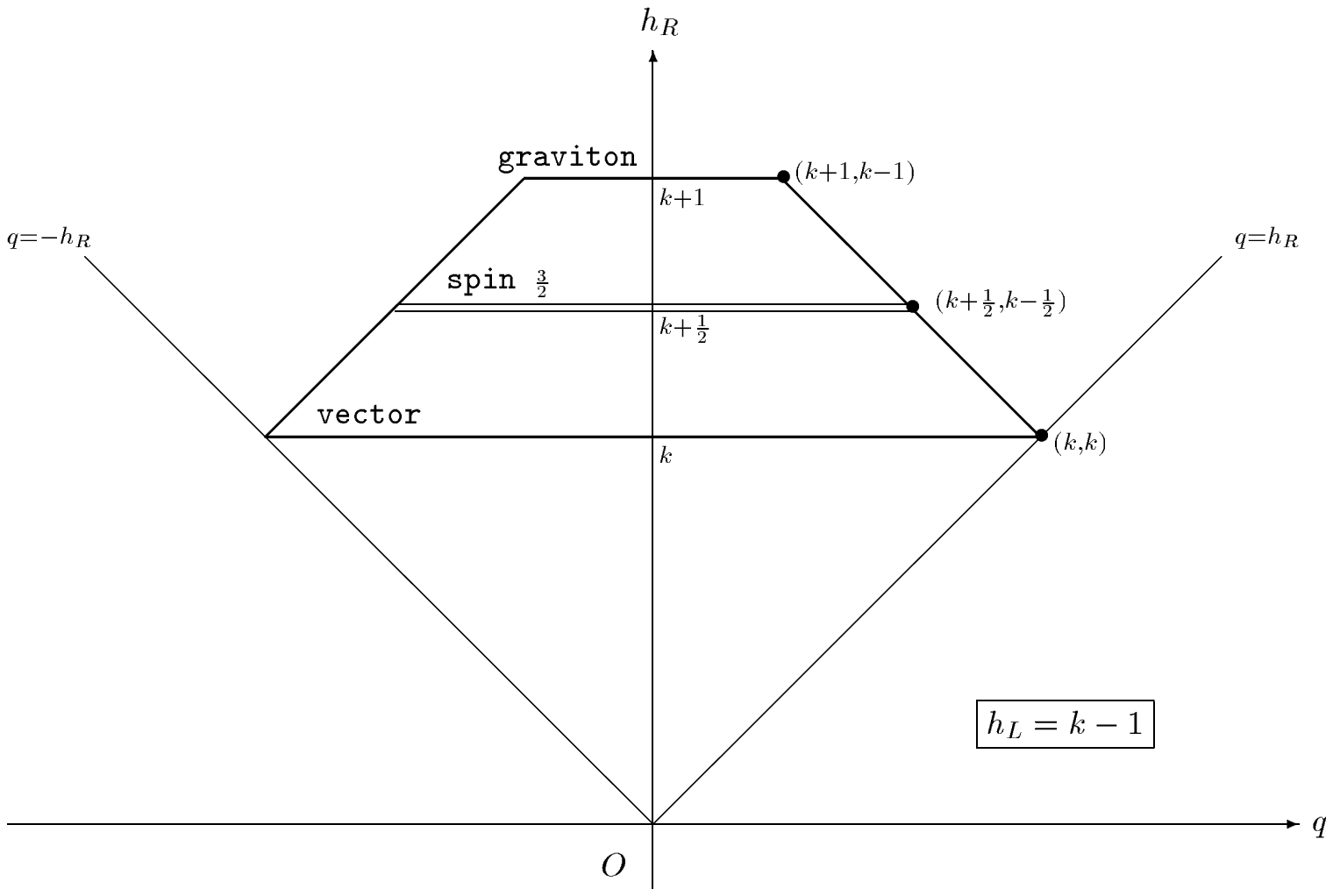}}
  \centerline{Fig.2(c)}
\end{figure}
\begin{figure}
  \epsfxsize = 12 cm   
  \centerline{\epsfbox{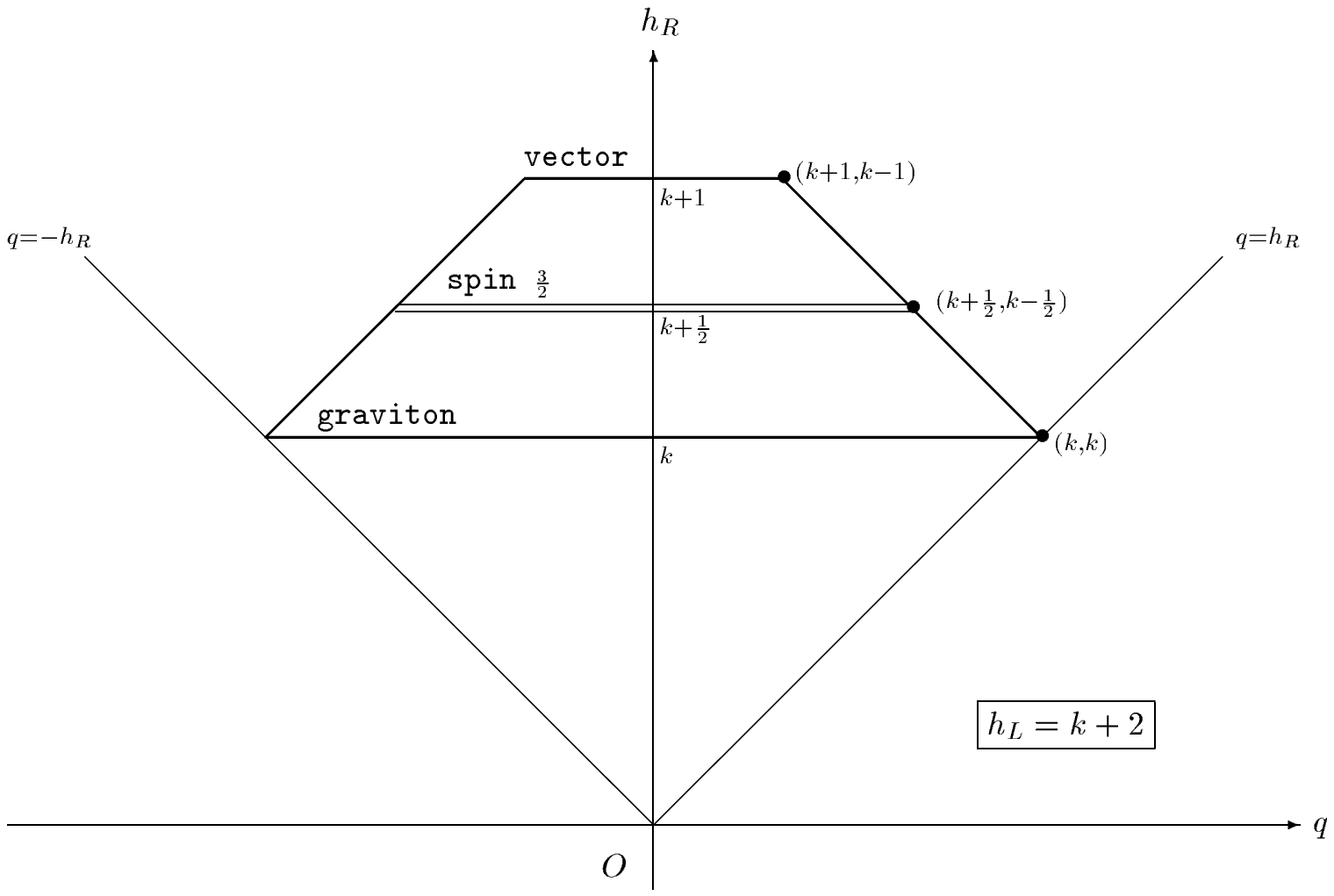}}
  \centerline{Fig.2(d)}
\end{figure}

\begin{thebibliography}{99}
%
\bibitem{Maldacena:1997re}
J.~Maldacena,
Adv.\ Theor.\ Math.\ Phys.\ {\bf 2} (1998) 231
hep-th/9711200.
%
\bibitem{Witten:1998qj}
E.~Witten,
Adv.\ Theor.\ Math.\ Phys.\ {\bf 2} (1998) 253
hep-th/9802150.
%
\bibitem{Gubser:1998bc}
S.S.~Gubser, I.R.~Klebanov and A.M.~Polyakov,
Phys.\ Lett.\ {\bf B428} (1998) 105
hep-th/9802109.
%
\bibitem{Aharony:1999ti}
O.~Aharony, S.S.~Gubser, J.~Maldacena, H.~Ooguri and Y.~Oz,
hep-th/9905111.
%
\bibitem{Gibbons:1995vm}
G.W.~Gibbons, G.T.~Horowitz and P.K.~Townsend,
Class.\ Quant.\ Grav.\ {\bf 12} (1995) 297
hep-th/9410073.
%
\bibitem{Mizoguchi:1998wv}
S.~Mizoguchi and N.~Ohta,
Phys.\ Lett.\ {\bf B441} (1998) 123
hep-th/9807111.
%
\bibitem{Balasubramanian:1998ee}
V.~Balasubramanian and F.~Larsen,
Nucl.\ Phys.\ {\bf B528} (1998) 229
hep-th /9802198.
%
\bibitem{Maldacena:1998bw}
J.~Maldacena and A.~Strominger,
JHEP {\bf 12} (1998) 005
hep-th/9804085.
%
\bibitem{Martinec:1998st}
E.J.~Martinec,
hep-th/9804111.
%
\bibitem{Fernando:1998eg}
S.~Fernando and F.~Mansouri,
hep-th/9804147.
%
\bibitem{Banados:1998pi}
M.~Banados, K.~Bautier, O.~Coussaert, M.~Henneaux and M.~Ortiz,
Phys.\ Rev.\ {\bf D58} (1998) 085020
hep-th/9805165.
%
\bibitem{BeBrGa}K. Behrndt, I. Brunner and I. Gaida,  
Phys. Lett. {\bf B432} (1998) 310; 
Nucl.\ Phys.\ {\bf B546} (1999) 65
hep-th/9806195.
%
\bibitem{Teo:1998dw}
E.~Teo,
Phys.\ Lett.\ {\bf B436} (1998) 269
hep-th/9805014.
%
\bibitem{Cvetic:1998xh}
M.~Cvetic and F.~Larsen,
Nucl.\ Phys.\ {\bf B531} (1998) 239
hep-th/9805097.
\bibitem{BaKrLa}V. Balasubramanian, P. Kraus and A. Lawrence, 
hep-th/9805171. 
%
\bibitem{Larsen:1998xm}
F.~Larsen,
Nucl.\ Phys.\ {\bf B536} (1998) 258
hep-th/9805208.
%
\bibitem{deBoer:1998ip}
J.~de Boer,
Nucl.\ Phys.\ {\bf B548} (1999) 139
hep-th/9806104.
%
\bibitem{Giveon:1998ns}
A.~Giveon, D.~Kutasov and N.~Seiberg,
Adv.\ Theor.\ Math.\ Phys.\ {\bf 2} (1998) 733
hep-th/9806194.
%
\bibitem{Rahmfeld:1998zn}
J.~Rahmfeld and A.~Rajaraman,
Phys.\ Rev.\ {\bf D60} (1999) 064014
hep-th/9809164.
%
\bibitem{Maldacena:1997de}
J.~Maldacena, A.~Strominger and E.~Witten,
JHEP {\bf 12} (1997) 002
hep-th/9711053.
%
\bibitem{Kallosh:1997qw}
R.~Kallosh and J.~Kumar,
Phys.\ Rev.\ {\bf D56} (1997) 4934
hep-th/9704189.
%
\bibitem{Boonstra:1997dy}
H.J.~Boonstra, B.~Peeters and K.~Skenderis,
Phys.\ Lett.\ {\bf B411} (1997) 59
hep-th/9706192.
%
\bibitem{Gunaydin:1986fe}
M.~Gunaydin, G.~Sierra and P.K.~Townsend,
Nucl.\ Phys.\ {\bf B274} (1986) 429.
%
\bibitem{Pilch:1984xy}
K.~Pilch, P.~van Nieuwenhuizen and P.K.~Townsend,
Nucl.\ Phys.\ {\bf B242} (1984) 377.
%
\bibitem{vanNieuwenhuizen:1985iz}
P.~van Nieuwenhuizen,
Class.\ Quant.\ Grav.\ {\bf 2} (1985) 1.
%
\bibitem{Gunaydin:1985wc}
M.~Gunaydin, P.~van Nieuwenhuizen and N.P.~Warner,
Nucl.\ Phys.\ {\bf B255} (1985) 63.
%
\bibitem{Dabholkar:1998kv}
A.~Dabholkar and J.A.~Harvey,
JHEP {\bf 02} (1999) 006
hep-th/9809122.
%
\bibitem{Dine:1997ji}
M.~Dine and E.~Silverstein,
hep-th/9712166.
%
\bibitem{Cr} E. Cremmer, in: ``{\it Cambridge 1980, Proceedings, 
Superspace and Supergravity}'', eds. S. W. Hawking
 and M. Roc\v{e}k (Cambridge University Press, 1981) 267.
%
\bibitem{Chamseddine:1980sp}
A.H.~Chamseddine and H.~Nicolai,
Phys.\ Lett.\ {\bf 96B} (1980) 89.
%
\bibitem{Biran:1984iy}
B.~Biran, A.~Casher, F.~Englert, M.~Rooman and P.~Spindel,
Phys.\ Lett.\ {\bf 134B} (1984) 179.
%
\bibitem{Freedman:1984na}
D.Z.~Freedman and H.~Nicolai,
Nucl.\ Phys.\ {\bf B237} (1984) 342.
%
\bibitem{Casher:1984ym}
A.~Casher, F.~Englert, H.~Nicolai and M.~Rooman,
Nucl.\ Phys.\ {\bf B243} (1984) 173.
%
\bibitem{EN}F. Englert and H. Nicolai, 
{\it ``Supergravity in Eleven-dimensional Space-time''},
12th Int. Colloq. on Group Theoretical Methods 
in Physics, Trieste Grp. Theor. Meth. (1983) 249.
%
\bibitem{Gunaydin:1986tc}
M.~Gunaydin and N.P.~Warner,
Nucl.\ Phys.\ {\bf B272} (1986) 99.
%
\bibitem{Duff:1986hr}
M.J.~Duff, B.E.~Nilsson and C.N.~Pope,
Phys.\ Rept.\ {\bf 130} (1986) 1.
%
\bibitem{Townsend:1984xs}
P.K.~Townsend, K.~Pilch and P.~van Nieuwenhuizen,
Phys.\ Lett.\ {\bf 136B} (1984) 38;
Addendum-ibid. \ {\bf 137B} (1984) 443. 
%
\bibitem{NiSeTa}
H.~Nicolai and E.~Sezgin,
Phys.\ Lett.\ {\bf 143B} (1984) 389.\\
%
H.~Nicolai, E.~Sezgin and Y.~Tanii,
Nucl.\ Phys.\ {\bf B305} (1988) 483.
%
\bibitem{Gibbons:1993sv}
G.W.~Gibbons and P.K.~Townsend,
Phys.\ Rev.\ Lett.\ {\bf 71} (1993) 3754
hep-th/9307049.
%
\bibitem{Freund:1980xh}
P.G.~Freund and M.A.~Rubin,
Phys.\ Lett.\ {\bf 97B} (1980) 233.
%
\bibitem{Regge:1957td}
T.~Regge and J.A.~Wheeler,
Phys.\ Rev.\ {\bf 108} (1957) 1063.
%
\bibitem{Kac}V.G.~Kac, Adv.\ in Math.\ {\bf 26} (1977) 8.
%
\bibitem{Vlad}
M.~Scheunert, W.~Nahm and V.~Rittenberg,
\JMP{17}{1976}{1626}; ibid.\ 1640.
%
\bibitem{Ademollo:1976wv}
M.~Ademollo {\it et al.},
Nucl.\ Phys.\ {\bf B114} (1976) 297.
%
\bibitem{Guenaydin}M. G{\"u}naydin, ``{\it Oscillator-like Unitary 
Representations of Non-Compact Groups and Supergroups and Extended 
Supergravity Theories}'' in the XI Colloquium on Group Theoretical 
Methods in Physics, Istanbul, 1982.
%
\bibitem{Brown:1986nw}
J.D.~Brown and M.~Henneaux,
Commun.\ Math.\ Phys.\ {\bf 104} (1986) 207.
%
\bibitem{Achucarro:1986vz}
A.~Achucarro and P.K.~Townsend,
Phys.\ Lett.\ {\bf B180} (1986) 89.
%
\bibitem{Townsend:1998ci}
P.K.~Townsend,
hep-th/9901102.
%
\bibitem{VanDerJeugt:1985hq}
J.~Van Der Jeugt,
J.\ Math.\ Phys.\ {\bf 26} (1985) 913.
%
\end{thebibliography}
\end{document}